\documentclass[journal,twoside,web]{ieeecolor}
\usepackage{tmi}
\usepackage{cite}
\usepackage{amsmath,amssymb,amsfonts}
\usepackage{algorithmic}
\usepackage{graphicx}
\usepackage{textcomp}
\usepackage{multirow}
\usepackage{subcaption}
\usepackage{placeins}  
\usepackage{hyperref}
\usepackage{float}
\usepackage{booktabs} 
\usepackage{array}

\begin{document}
\bstctlcite{IEEEexample:BSTcontrol}
\title{AI-Assisted Colonoscopy: Polyp Detection and Segmentation using Foundation Models}
\author{
Uxue Delaquintana-Aramendi*,
Leire Benito-del-Valle*,
Aitor Alvarez-Gila,
Javier Pascau,
Luisa F Sánchez-Peralta,
Artzai Picón,
J Blas Pagador,
Cristina L Saratxaga
\thanks{Submited 2025/03/27.}
\thanks{* Equal contribution.}
\thanks{This work was supported in part by the Elkartek Programme, Basque Government (Spain) (BEREZ-IA, KK-2023/00012 and IKUN, KK-2024/00064).}
\thanks{U. Delaquintana-Aramendi was with TECNALIA, Basque Research and Technology Alliance (BRTA), Derio, Bizkaia, 48160 Spain and with Universidad Carlos III de Madrid, Madrid, 28911 Spain. She is now with Artificial Intelligence in Science Institute (AISI) at University of California Irvine, CA 92617 USA (e-mail: udelaqui@uci.edu).}
\thanks{L. Benito-del-Valle is with TECNALIA, Basque Research and Technology Alliance (BRTA), Derio, Bizkaia, 48160 Spain (e-mail: leire.benitodelvalle@tecnalia.com).}
\thanks{A. Alvarez-Gila is with TECNALIA, Basque Research and Technology Alliance (BRTA), Derio, Bizkaia, 48160 Spain (e-mail: aitor.alvarez@tecnalia.com).}
\thanks{J. Pascau is with the Department of Bioengineering at Universidad Carlos III de Madrid, Madrid, 28911 Spain and with Instituto de Investigación Sanitaria Gregorio Marañón, Madrid, 28009 Spain (e-mail: jpascau@ing.uc3m.es).}
\thanks{L. F Sánchez-Peralta is with Jesús Usón Minimally Invasive Surgery Centre, Cáceres, 10071 Spain and AI4polypNET Thematic Network, Barcelona, 08193 Spain (e-mail: lfsanchez@ccmijesususon.com).}
\thanks{A. Picón is with TECNALIA, Basque Research and Technology Alliance (BRTA), Derio, Bizkaia, 48160 Spain and University of the Basque Country, Plaza Torres Quevedo, 48013 Bilbao, Spain (e-mail: artzai.picon@tecnalia.com).}
\thanks{J. Blas Pagador is with Jesús Usón Minimally Invasive Surgery Centre, Cáceres, 10071 Spain and AI4polypNET Thematic Network, Barcelona, 08193 Spain (e-mail: jbpagador@ccmijesususon.com).}
\thanks{C. L Saratxaga is with TECNALIA, Basque Research and Technology Alliance (BRTA), Derio, Bizkaia, 48160 Spain (e-mail: Cristina.Lopez@tecnalia.com).}
}

\IEEEaftertitletext{\vspace{-3\baselineskip}}
\maketitle

\begin{abstract}
In colonoscopy, 80\% of the missed polyps could be detected with the help of Deep Learning models. In the search for algorithms capable of addressing this challenge, foundation models emerge as promising candidates. Their zero-shot or few-shot learning capabilities, facilitate generalization to new data or tasks without extensive fine-tuning. A concept that is particularly advantageous in the medical imaging domain, where large annotated datasets for traditional training are scarce. In this context, a comprehensive evaluation of foundation models for polyp segmentation was conducted, assessing both detection and delimitation. For the study, three different colonoscopy datasets have been employed to compare the performance of five different foundation models, DINOv2, YOLO-World, GroundingDINO, SAM and MedSAM, against two benchmark networks, YOLOv8 and Mask R-CNN. Results show that the success of foundation models in polyp characterization is highly dependent on domain specialization. For optimal performance in medical applications, domain-specific models are essential, and generic models require fine-tuning to achieve effective results. Through this specialization, foundation models demonstrated superior performance compared to state-of-the-art detection and segmentation models, with some models even excelling in zero-shot evaluation; outperforming fine-tuned models on unseen data.\looseness=-1
\end{abstract}

\begin{IEEEkeywords}
Colonoscopy, Polyp Detection, Polyp Segmentation, Deep Learning, Foundation Models.

\end{IEEEkeywords}

\section{Introduction}
\label{sec:introduction}
\IEEEPARstart{C}{olorectal cancer} (CRC) is the third most diagnosed cancer worldwide, accounting for 10\% of all cancers cases~\cite{CRC}. Only in the USA, the American Cancer Society (ACS) estimates about 107,320 new cases of colon cancer for 2025~\cite{acs_colorectal_2025}. In terms of mortality, colorectal cancer is the second leading cause of cancer death, only behind lung cancer~\cite{ccstats}. As such, ACS expects it to cause about 52,900 deaths for 2025. Despite this, cure rates can reach 90\% if patients are diagnosed in the early stages of the disease~\cite{cure}.

CRC originates from the inner lining of the colon or rectum~\cite{origin}, where abnormal growths of tissue known as colon polyps, can arise due to the unidentified accumulation of genetic mutations. These polyps can progress from a benign to a carcinogenic state over time~\cite{malignantpolyp}.  

Colonoscopy is the standard technique used for detecting and removing these anomalies; a vital tool in the early detection and prevention of CRC~\cite{piccoloproject}. Some polyps are easily identified and removed directly for examination. However, there are several factors that complicate their detection. Polyps often appear close to other anatomical structures, such as folds or blood vessels, which can cause confusion when defining their shape. Furthermore, polyps exhibit inconsistent morphology, with certain shapes posing particular challenges: flat or small polyps are notoriously difficult for the human eye to detect~\cite{missedpolyps}. The quality of colonoscopy videos also varies significantly; issues related to the image acquisition tool and inadequate bowel preparation prior to the procedure can further obscure the visibility. As a result, it has been determined that 30\% of polyps remain undetected after a colonoscopy~\cite{SOTA1}. 

Given this significant gap, there is growing optimism that recent advances in computer vision through Deep Learning (DL) models could lead to the development of effective algorithms to assist clinicians in the task and improve prognosis. 

In this sense, \textbf{foundation models} have emerged as promising candidates due to their strong zero-shot and few-shot generalization abilities, achieved through pretraining on massive amounts of data~\cite{foundationmodels}. The potential of these DL models lies in the ability to perform successfully on tasks or data they have never encountered before, with little, few-shot learning, or no, zero-shot learning, fine-tuning on task-specific data. In the medical domain, where large, annotated datasets for traditional model training are scarce, these zero-shot and few-shot learning approaches can be particularly advantageous. 

Foundation models can stem from different strategies. Some rely on the principle that training on large amounts of data leads to a broader range of references and improved generalization. Promptable segmentation models like Meta's Segment Anything Model (SAM)~\cite{SAM}, pretrained on 11 million annotated natural images, or MedSAM~\cite{MedSAM} (SAM fine-tuned on 1.09 million medical image-mask pairs), exemplify this approach. On a similar note, self-supervised DINOv2~\cite{dinov2} (self-DIstillation with NO labels), trained with 142 million images, provides high-performance features and can be used to create multipurpose backbones. It has shown strong prediction capabilities on tasks ranging from classification or segmentation to image retrieval. In other cases, a traditional model's understanding can be enhanced by integrating these powerful architectures. In this regard, YOLO-World~\cite{yoloworld} and GroundingDINO (GDINO)~\cite{groundingdino_paper} are classic object detectors augmented with the ability to interpret text by adding foundation models specialized in relating text to images.\looseness=-1 

Although efforts have been made to adapt foundation models to the medical domain~\cite{conference}, an evaluation of state-of-the-art foundation models for polyp detection and segmentation remains elusive. The domain shift between the data used to pretrain these models and the specific requirements of medical tasks underscores the need for targeted assessment to benchmark performance, identify improvement gaps, compare zero-shot (pretrained model) and few-shot (fine-tuned model) capabilities, and overall, quantify the effectiveness of these models within this specialized setting. 

In this context, this study aims to develop a pipeline capable of accurately detecting and segmenting polyps in colonoscopy. The objective is twofold: first, to assess the effectiveness of foundation models in the colonoscopy segmentation domain; and, if possible, to obtain a foundation model-based segmentation model that outperforms traditional methods with minimal or no training on domain-specific data. 

To achieve this, both original and fine-tuned versions of the selected models were evaluated against traditional DL approaches in an exhaustive benchmarking process. The models were trained and tested on three distinct colonoscopy datasets to replicate the diverse scenarios inherent in clinical practice, thus ensuring the robustness and validity of the results.\looseness=-1 

\section{Related Work}
Advances in DL revolutionized polyp detection and segmentation in colonoscopy, with Convolutional Neural Networks (CNNs) at the forefront of this progress~\cite{miccai}. CNN-based models like VGG and ResNet, have become pivotal in this field due to their ability to extract hierarchical feature representations, as demonstrated by Urban et al.~\cite{urban2018}, who achieved 96\% accuracy in real-time polyp detection and localization. Additionally, Region-based CNNs, including Mask R-CNN~\cite{maskrcnn}, have been widely adapted for precise segmentation, leveraging selective search techniques to improve identification and classification.\looseness=-1 

It is known that encoder-decoder architectures, particularly U-Net and its derivatives, have seen extensive use in medical image segmentation. A good example is the work by Chen et al.~\cite{chenbaldi}, where detailed pixel-level segmentation of polyps could be obtained. More recently, attention mechanisms have further advanced the field, with models like SR-AttNet~\cite{alamfattah} improving segmentation by focusing on relevant features. Additionally, transformer-based architectures have demonstrated state-of-the-art performance, with models such as 3D TransUNet effectively capturing global context and enhancing accuracy~\cite{chen}.

In any case, the effectiveness of these models often depends on access to large, high-quality labeled datasets, which are difficult and costly to obtain in the biomedical field. Lately, foundation models have shown promise in generalizing to new tasks without the need for task-specific fine-tuning (zero-shot learning). The application of foundation models in polyp segmentation is still in its early stages; however, Meta's SAM has already seen widespread adoption for medical imaging, with models like SAM-EG~\cite{sameg}, Polyp-SAM++~\cite{polypsam}, and the aforementioned MedSAM~\cite{MedSAM}. Notably, MedSAM exhibits an impressive accuracy of 98.5\% in polyp segmentation.\looseness=-1

Nonetheless, it is important to note that the success of such models is highly dependent on the quality of the task-specific prompts provided alongside the image. This suggests that proper segmentation is difficult to achieve without accurate prior detection.\looseness=-1 

\section{Method}
We frame the polyp detection problem as an instance segmentation task, and address it by combining the detection of polyps, if present, and their subsequent segmentation to accurately characterize their morphology. This approach involves the use of two models: the first model detects the presence of polyps and, if any, generates a bounding box around each one. These box prompts are then passed to the segmentation model, responsible for producing the final shape mask.

\subsection{Object Detection Models}
Three object detection models were evaluated: one baseline or reference model, \textbf{YOLOv8}~\cite{yolo}, and two foundation models, \textbf{YOLO-World}~\cite{yoloworld} and \textbf{GroundingDINO}~\cite{groundingdino_paper}. 

\textbf{YOLOv8} is the eight version of the YOLO (\textit{You Only Look Once}) algorithm series, a well-known family of object detection and classification models in computer vision.

\textbf{YOLO-World} can be regarded as the first multimodal foundation model of the YOLO family.
It extends YOLOv8 with open-vocabulary detection capabilities through the addition of a text encoder based on CLIP~\cite{clip}, a foundation model that uses contrastive learning to align image and text embeddings.
Thus, it gains the ability to operate in an open setup, recognizing and locating objects in images, even if the specific prompt-defined target categories were not part of the training dataset.
In this study, the pre-trained \textit{YOLOv8s-Worldv2} model from Ultralytics is employed.

Following the same principle, Grounding DINO (\textbf{GDINO}), extends DINO~\cite{dino} (\textit{DETR with Improved Denoising Anchor Boxes}) to open-set object detection, enabling it to detect arbitrary objects given text as prompts.
This is achieved by combining DINO with BERT~\cite{bert}.
Specifically, we evaluated GDINO with the Swin-T backbone~\cite{groundingdino_paper}. 

\subsection{Segmentation Models}
For the segmentation portion of the study, two foundation models, \textbf{SAM}~\cite{SAM} and \textbf{MedSAM}~\cite{MedSAM}, were compared against a strong baseline, \textbf{Mask R-CNN}~\cite{maskrcnn}.

Mask R-CNN (\textbf{MR-CNN}) enables instance segmentation by extending the Faster R-CNN object detection model through a ``mask head" that creates the pixel-wise segmentation mask for each detected object. The architecture combines a backbone network followed by a Feature Pyramid Network (FPN), which acts as a neck to fuse the multi-level features extracted by the backbone.
In this study, the model was tested with two different backbones for feature extraction: ResNet-FPN-50 (\textbf{R50}), proposed in the original architecture, and \textbf{DINOv2} in its smallest version~\cite{dinov2}, as a novel approach.
Additionally, given that MR-CNN can yield bounding boxes as well, it was included as an additional baseline for object detection, together with YOLOv8.

\textbf{SAM} is a foundation model specialized in promptable image segmentation. It delivers state-of-the-art results for interactive, flexible and class-agnostic segmentation across domains, supporting different types of input prompts.
In this case, bounding box prompts were combined with the input image. The smallest version of SAM, \textit{}{SAM Base}, was chosen for evaluation; as MedSAM’s fine-tuning stems from it.

Lastly, \textbf{MedSAM} represents a fine-tuned version of SAM, and was introduced as the first foundation model for universal medical image segmentation.
Like SAM, MedSAM uses prompts (bounding boxes) along with the input image to generate the segmentation mask. MedSAM builds on SAM’s architectural structure and it is trained with a medical image dataset composed of 1,090,486 annotated image-mask pairs.
The dataset reflects 15 imaging modalities, more than 30 cancer types, and many imaging protocols.

\subsection{Datasets}
Three colonoscopy datasets, all in the same format, were selected for fine-tuning and evaluating the models:

The \textbf{PICCOLO}~\cite{PICCOLO} dataset comprises 3,433 annotated frames that showcase 76 different polyps, accompanied by clinical metadata, where the details of each polyp are specified: size (in mm), Paris classification~\cite{Paris}, NICE classification~\cite{nice} and preliminary diagnosis. 
The dataset features images captured using both white-light (WL) and narrow-band imaging (NBI) technologies: 2,131 WL images and 1,302 NBI images. Each frame was manually annotated to create the binary polyp masks. The representation of negative images (no polyp shown) is as follows; 0.91\% in the train set, 3.12\% in validation and 0.3\% in test. 

\textbf{PolypSegm-ASH}~\cite{polypsegm-ash} is a novel dataset containing 1,226 high-definition images with a total of 473 unique polyps.
Images were pixel-wise annotated by clinicians creating the polyp masks. No negative images are reported to be included.
Additionally, PolypSegm-ASH is part of the iSMIT 2024 Polyp Segmentation Challenge. As such, at the time of access for this study, annotations for the test set were not publicly available.\looseness=-1

Lastly, the \textbf{SUN-SEG}~\cite{sunseg} dataset comprises 158,690 colonoscopy images, of which 49,136 contain annotated polyps. There are 100 positive cases, each containing images of a unique polyp, which were split into training (70\%), validation (20\%), and testing (10\%) sets, resulting in 37,999 training, 8,865 validation, and 2,272 testing images. Negative images, derived from 13 different patient videos, were divided similarly, with approximately 9 cases for training, 3 for validation, and 1 for testing. To select the same amount of negative images per set and ensure balanced representation, a calculated "temporal jump" was used to space the selected images, maintaining dataset variability. A total of 98,270 (49,136 positive, 49,136 negative) images were used for this study.\looseness=-1

Note that the training parameters, evaluation prompts, and overall algorithm were specifically designed and optimized using the PICCOLO dataset. To validate our approach and demonstrate its generalization capabilities, the final version of each model was then tested on the PolypSegm-ASH and SUN-SEG datasets. 

\subsection{Algorithm Setup}
\label{algorithmsetup}

\begin{figure}[t]
    \centering
    \includegraphics[width=\linewidth]{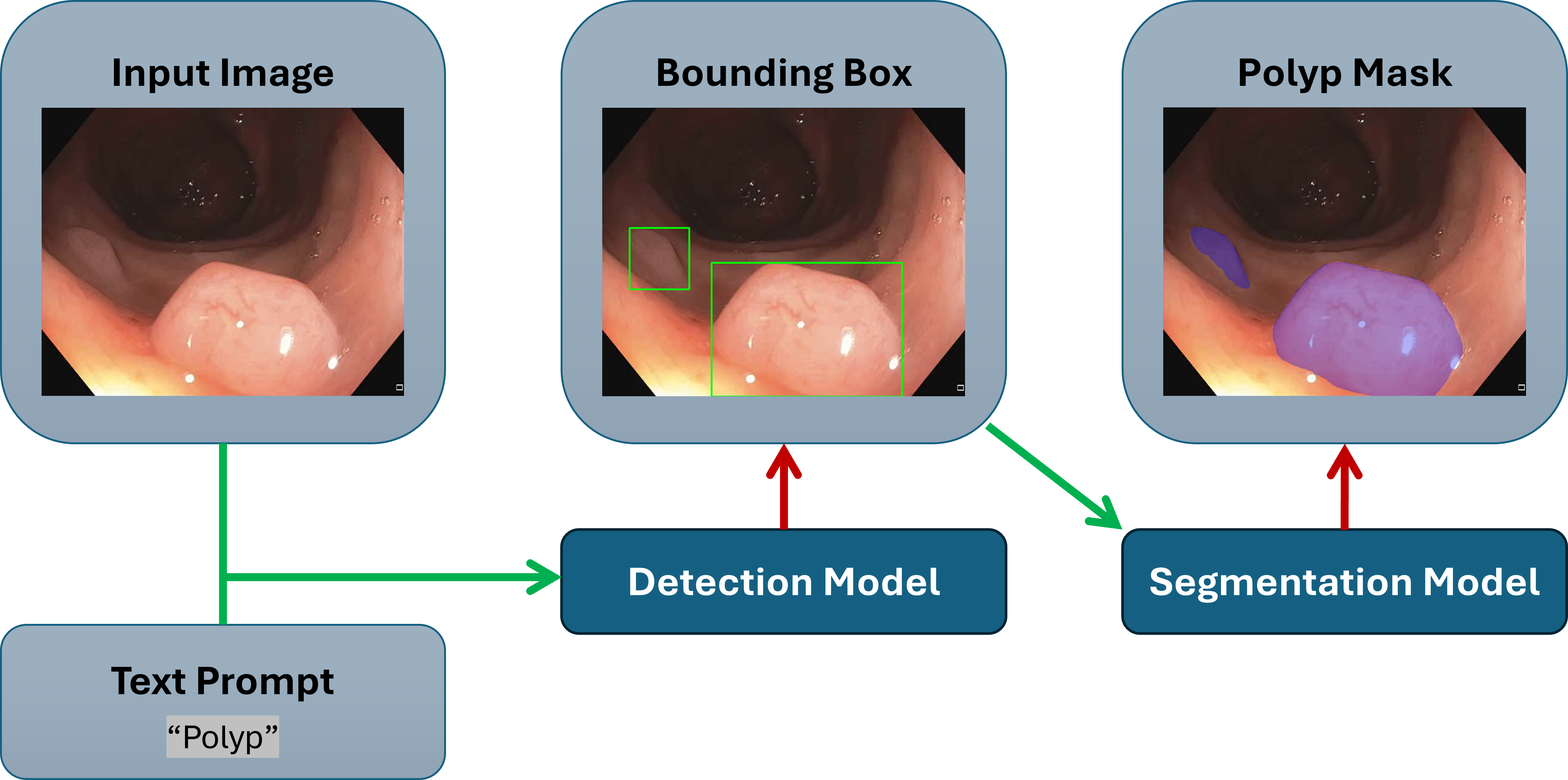}
    \caption{\textbf{Algorithm workflow diagram.} The process begins with an image input, accompanied by a text prompt (optional) specifying the object of interest. A detection model generates bounding boxes around detected objects, which are then passed, along with the original image, to a segmentation model. The latter produces the final polyp mask.}
    \label{fig:graph}
\end{figure}

To build the algorithm pipeline, the detection model is linked to the segmentation model (see Fig.~\ref{fig:graph}), except for the case of MR-CNN, where the entire segmentation process is performed atomically. 

The workflow begins by feeding the images in the correct format to each of the detection models under study. For the few-shot models (fine-tuned), the test images were resized to the dimensions of the corresponding training images (see Section~\ref{Training}), to maintain consistency and ensure optimal results. For zero-shot YOLO-World, test images were resized to the default dimension of 640x640, which was maintained across all three datasets. Zero-shot GDINO was evaluated with the original data dimensions, as the model is not square-centered. Inference on square images was tested aiming for better results, but no improvement was observed.\looseness=-1 

Apart from the images, YOLO-World and GDINO also require a text query as input, where the target object class is specified. Several context-specific prompts were tested: ``polyp", ``tumor" and ``lump". In GDINO ``lump" proved to be the most effective prompt for evaluation; with an Average Recall value of 0.52, against 0.446 for ``tumor" and 0.427 for ``polyp". Whereas, in the case of YOLO-World, none of the above succeeded in predicting any bounding boxes. Therefore, inference was run without any text prompt to access the model’s intrinsic natural language, which yielded proper polyp detection. Fine-tuned models didn't require any further specifications, as the prompt ``polyp" was hard-coded in training. 
Additionally, a minimum confidence value of 0.2 was set for GDINO detections to be meaningful. While confidence levels around 0.5 are typically preferred, lower thresholds were necessary to achieve narrower detections in this context.

From an exploration on the performance of SAM and MedSAM for polyp segmentation using PICCOLO's validation set, it was concluded that adding a fixed augmentation of 10 pixels per side to the bounding box prompt enhanced the segmentation, reaching an accuracy of almost 93\%, which aligns with the 98.5\% accuracy reported by the authors~\cite{MedSAM} on the task. The authors did not specify the design of bounding boxes for testing that yielded such high values.
As such, various approaches had to be tested, and ground truth-based augmentation provided the best results.
This impressive zero-shot performance discarded the need for fine-tuning SAM and MedSAM on our target datasets, as it became evident that this could only bring marginal improvements to the overall performance. 

In order to ensure compatibility, images were resized to $1024\times1024$ before feeding them to the segmentation model, and the pixel values were normalized to match the expected intensity range [0,255]. The same process was applied to the bounding box prompts. Finally, SAM and MedSAM yield the predicted segmentation mask within the input bounding box, thus obtaining one binary segmentation mask for each instance of detected polyp.

\subsection{Training}
\label{Training}
For the baseline models we directly evaluated their fine-tuned version, while, for the foundation models, we considered both the zero-shot (non-fine-tuned) and few-shot (fine-tuned) performances, except for MedSAM and SAM, due to the reasons outlined in Section~\ref{algorithmsetup}.

The datasets were split in train, validation and test subsets. We used 2,203 frames for train, 897 for validation, and 333 frames for test in PICCOLO; 75,996 train, 17,730 validation and 4,544 test frames in SUN-SEG. In the case of the PolypSegm-ASH dataset, as no annotations were available for the test set, the same frames were used for validation and test; yielding 788 train and 113 validation/test frames. 

For fine-tuning on YOLOv8 and YOLO-World, the specifications found in the Ultralytics YOLO repository were followed. YOLO-based models work with square images and resize them to $640\times640$ by default for inference. Yet, in training, individual values were chosen for the three different datasets in use, aiming for proximity to our original image resolution and for the least deformation possible. The resulting resolutions were $608\times608$ for PICCOLO ($600\times480$ originally), $1152\times1152$ for SUN-SEG ($1240\times1080$ originally) and $1440\times1440$ for PolypSegm-ASH ($1280\times1024$ originally). Both models were trained for 100 epochs on the default parameters~\cite{yolov8_doc, yoloworld_doc}. 

For fine-tuning MR-CNN, the model hyperparameters were optimized over PICCOLO dataset's validation set the with a particular focus on the learning rate (LR). 
An analysis of the performance of different initial LR values determined that 0.00001 yielded the best results, based on the average precision and mean IoU on the validation set. Consistent with previous training, the training split of each dataset was resized to fit the model requirements: $600\times600$ for PICCOLO, $800\times800$ for SUN-SEG, and $1440\times1440$ for PolypSegm-ASH.

Significant effort was directed towards optimizing the MR-CNN parameters to establish a robust baseline model, enabling a competitive basis for comparison. In contrast, the foundation models were employed without parameter optimization to align with their primary advantage of minimizing the time and effort required for task-specific training.

In addition to the original MR-CNN backbone for feature extraction (i.e. ResNet-50-FPN), DINOv2-FPN was implemented as an alternative, keeping the FPN~\cite{dinov2}. The same training procedure was used for both backbones, but parameter optimization was conducted on the original model before being applied to the DINOv2-FPN architecture. This likely gives ResNet-50-FPN an advantage, however, it aligns with our goal of minimal adaptation for foundation models.

Lastly, fine-tuning on GDINO was based on the implementation by MMDetection. Swin-T backbone was used for single-class (``polyp") detection, training during 100 epochs and an initial learning rate of 0.00005, which was then modified with warm-up learning rate scheduling and multi-step decay during training. The images were used in their original format and the rest of the training parameters were set to default~\cite{groundingdino_mmdetection_ft}. 

\subsection{Evaluation scheme}
The evaluation was divided in two parts: a preliminary evaluation mid-way, where the object detection models are assessed, and a segmentation evaluation of the final masks, conducted at the end of the pipeline. 

\subsubsection{Object Detection Evaluation}

The bounding boxes predicted by YOLOv8, YOLO-World,
GDINO and MR-CNN, were evaluated against the ground truth annotations using Average Precision (AP) and Average Recall (AR) object detection metrics, as computed in Python COCO Tools~\cite{coco}.

In this context, AP evaluates the overall effectiveness of polyp detection, while AR accounts for false negatives; making it particularly valuable in the context of colonoscopy, where missing a polyp detection can have serious consequences. AP and AR were averaged over multiple Intersection over Union (IoU) threshold values. Specifically, 10 IoU thresholds were used from 0.50 to 0.95, with a step size of 0.05. The IoU is the measure of overlap between the original and predicted boxes, and, in this case, it is used as the threshold above which detections are considered true positives. Averaging over IoUs rewards detectors with better localization~\cite{coco}. 

\subsubsection{Segmentation Evaluation}

Segmentation models are evaluated using Intersection over Union (IoU), which measures the overlap between predicted masks and ground truth masks.

The algorithm handles two critical challenges: missed detections and accurate predictions of negative frames. Ground truth masks are binarized for comparison, and IoU is calculated for each frame. If no bounding boxes are predicted and the ground truth contains no significant object (fewer than 20 white pixels), the model correctly predicts a negative frame (IoU = 1). Otherwise, missed detections are penalized based on the ground truth. When bounding boxes are present, predicted objects are combined into a single binarized mask, and IoU is calculated against the ground truth. The final evaluation is the mean IoU across all test frames.  

\subsubsection{Conditional evaluation}
\label{conditional_evaluation}

Apart from the general evaluation explained above, the study utilized the metadata from the PICCOLO dataset to conduct a conditional evaluation, structured around four main categories designed to assess the impact of various factors on polyp detection performance. The first category examined the type of light used to capture the frame, differentiating between white light (WL) and narrow-band imaging (NBI). The second category followed the NICE classification, which categorizes polyps according to their carcinogenic potential, ranging from type 1 (least developed) to type 3 (most likely to be cancerous). The third category assessed polyps by size, categorizing them as diminutive or small, while the fourth category classified polyps based on their shape, according to the Paris classification, which distinguishes between protruded, flat elevated, and flat polyps.

Independently, the robustness of the models to the different lighting conditions mentioned was examined, as a factor with important clinical implications. While NBI is known to enhance polyp visibility, it is less accessible than WL and is typically used only after a potential anomaly is identified using WL. Therefore, a model that performs well under both light conditions, or even better under WL, would be highly advantageous in clinical practice.\looseness=-1

\section{Results}

\subsection{General Results}

The detection performance metrics, including Average Precision (AP) and Average Recall (AR), across three datasets (PICCOLO, PolypSegm-ASH, and SUN-SEG), are summarized in Table~\ref{detection_table}.

\begin{table}[tpb!]  
\centering
\caption{\textbf{Detection Performance (AP, AR)}. Baseline (top block) and foundation model-based methods (bottom block).}
\label{detection_table}
\setlength{\tabcolsep}{3pt}
\resizebox{\columnwidth}{!}{ 
\begin{tabular}{lccccccc}
\toprule
\multirow{2}[3]{*}{MODEL} & \multirow{2}[3]{*}{FT} & \multicolumn{2}{c}{PICCOLO} & \multicolumn{2}{c}{PolypSegm-ASH} & \multicolumn{2}{c}{SUN-SEG} \\
\cmidrule(lr){3-4} \cmidrule(lr){5-6} \cmidrule(lr){7-8}

                       &                     & AP & AR                       & AP & AR                              & AP & AR                       \\
\midrule
Yolov8                 & \checkmark          & 0.529 & 0.587                      & 0.588 & 0.655                             & 0.471 & 0.494                      \\ 
MR-CNN R50   & \checkmark          & 0.571 & 0.614                      & 0.705 & 0.761                             & 0.472 & 0.507                      \\ 
\midrule
MR-CNN DINOv2      & \checkmark          & 0.438 & 0.486                      & 0.348 & 0.396                             & 0.311 & 0.352                      \\ 
YOLOWorld             &                     & 0.081 & 0.159                      & 0.055 & 0.226                             & $<0.001$ & 0.006                      \\ 
YOLOWorld             & \checkmark          & 0.435 & 0.495                      & 0.613 & 0.681                             & 0.497 & 0.524                      \\ 
GDINO         &                     & 0.133 & 0.520                      & 0.072 & 0.369                             & 0.002 & 0.086                      \\ 
GDINO         & \checkmark       & \textbf{0.722} & \textbf{0.805}                     & \textbf{0.794} & \textbf{0.846}                             & \textbf{0.514} & \textbf{0.547}\\
\bottomrule
\end{tabular}
} 
\end{table}

Among the evaluated models, fine-tuned GDINO consistently achieved the highest scores across all datasets, with AP and AR values of 0.722 and 0.805 for PICCOLO, 0.794 and 0.846 for PolypSegm-ASH, and 0.514 and 0.547 for SUN-SEG. MR-CNN R50 with fine-tuning also demonstrated competitive performance, particularly on the PolypSegm-ASH dataset, where it attained an AP of 0.705 and an AR of 0.761. Significant improvement in performance was observed across all models after fine-tuning; for example, YOLO-World improved on PICCOLO from an AP of 0.081 and AR of 0.159 to 0.435 and 0.495, respectively. Notably, the baseline versions of the foundation models exhibited extremely poor performance, with exceptionally low precision and recall values for YOLO-World in particular: $<0.001$ and 0.006, respectively.

The segmentation performance achieved after combining the detection model with SAM, MedSAM and MR-CNN is shown in Table \ref{segmentation_table}. These results underscore the crucial role of detection quality, as the best segmentation performance corresponds to the most effective detection model. Fine-tuned GDINO paired with MedSAM achieved the highest segmentation scores across all datasets, obtaining values of 0.789 for PICCOLO, 0.885 for PolypSegm-ASH, and 0.799 for SUN-SEG. Fine-tuned MR-CNN R50 paired with MedSAM followed, with scores of 0.687, 0.835 and 0.763, respectively. The MedSAM configuration outperformed SAM in the vast majority of cases, surpassing even the strong segmentation capabilities of integrated detection-segmentation models like MR-CNN R50 and MR-CNN DINOv2, which achieved higher scores than SAM but still fell short of MedSAM's performance.\looseness=-1

\begin{table}[tpb!]  
\centering
\caption{\textbf{Segmentation Performance (mIoU).} Baseline (top) and foundation model-based methods (bottom block). FT=Fine-Tuned.}
\label{segmentation_table}
\setlength{\tabcolsep}{2pt}
\resizebox{\columnwidth}{!}{ 
\begin{tabular}{lclcccc}
\toprule

\multicolumn{2}{c}{Detection} & \multicolumn{2}{c}{Segmentation} & \multirow{2}[3]{*}{PICCOLO} & \multirow{2}[-1]{*}{PolypSegm} & \multirow{2}[3]{*}{SUN-SEG} \\
\cmidrule(lr){1-2} \cmidrule(lr){3-4}
MODEL & FT & MODEL & FT & & -ASH & \\

\midrule
MR-CNN R50 & \checkmark & MR-CNN R50 & \checkmark & 0.681 & 0.829 & 0.764 \\
\midrule
MR-CNN R50 & \checkmark & SAM & & 0.655 & 0.754 & 0.717 \\
MR-CNN R50 & \checkmark & MedSAM & &  0.687 & 0.835 &  0.763\\ 
MR-CNN DINOv2 & \checkmark & MR-CNN DINOv2 & \checkmark & 0.632 &  0.578 & 0.722 \\
MR-CNN DINOv2 & \checkmark & SAM & & 0.649 & 0.623 & 0.701 \\
MR-CNN DINOv2 & \checkmark & MedSAM & & 0.648 & 0.656 & 0.744 \\ 
Yolov8 & \checkmark & SAM & & 0.628 & 0.701 & 0.727 \\
Yolov8 & \checkmark & MedSAM & & 0.667 & 0.762 & 0.752 \\ 
YOLOWorld &         & SAM & &  0.175 & 0.163 & 0.351 \\
YOLOWorld &         & MedSAM & & 0.187 & 0.193 & 0.352 \\ 
YOLOWorld & \checkmark & SAM & & 0.555 & 0.703 & 0.708 \\
YOLOWorld & \checkmark & MedSAM & & 0.584 & 0.747 & 0.769 \\ 
GDINO &    & SAM & & 0.088 & 0.146 & 0.02 \\
GDINO &    & MedSAM & & 0.239 & 0.178 & 0.02 \\
GDINO & \checkmark & SAM & & 0.703 & 0.813 & 0.749 \\
GDINO & \checkmark & MedSAM & & \textbf{0.789} & \textbf{0.885} & \textbf{0.799} \\
\bottomrule
\end{tabular}
} 
\end{table}

\subsection{Conditional Results}

\begin{table*}[tpb!]  
\centering
\caption{Detection Conditional Evaluation}
\label{conditional_eval_table}
\setlength{\tabcolsep}{3pt}
\begin{tabular}{lcccccccccccccccccccccc}
\toprule
\multirow{3}[3]{*}{MODEL} & \multirow{3}[3]{*}{FT} & \multicolumn{4}{c}{Light Type} & \multicolumn{6}{c}{NICE Type} & \multicolumn{4}{c}{Polyp Size} & \multicolumn{6}{c}{Paris Classification} \\ 
\cmidrule(lr){3-6} \cmidrule(lr){7-12} \cmidrule(lr){13-16} \cmidrule(lr){17-22}
                       &                     & \multicolumn{2}{c}{WL} & \multicolumn{2}{c}{NBI} & \multicolumn{2}{c}{1} & \multicolumn{2}{c}{2} & \multicolumn{2}{c}{3} & \multicolumn{2}{c}{Diminutive} & \multicolumn{2}{c}{Small} & \multicolumn{2}{c}{Protruded} & \multicolumn{2}{c}{Flat Elevated} & \multicolumn{2}{c}{Flat} \\ 
\cmidrule(lr){3-4} \cmidrule(lr){5-6} \cmidrule(lr){7-8} \cmidrule(lr){9-10} \cmidrule(lr){11-12} \cmidrule(lr){13-14} \cmidrule(lr){15-16} \cmidrule(lr){17-18} \cmidrule(lr){19-20} \cmidrule(lr){21-22} 
                       &                     & AP & AR & AP & AR & AP & AR & AP & AR & AP & AR & AP & AR & AP & AR & AP & AR & AP & AR& AP & AR\\ 
\midrule
Yolov8                 & \checkmark          &  0.51     &   0.56    &       0.55     &  0.63  &   0.44   &   0.47    &  0.66     &  0.73     &   0.52    &   0.59 &    0.50   &    0.54   &  0.65     & 0.76 &   0.75    &   0.80    &  0.59     &  0.64     &  0.15     &  0.21   \\ 
YOLOWorld           &   &    0.08                 &   0.20    &   0.07    &         0.10    &    0.01    & 0.04      & 0.08      &  0.08     &  0.15           &  0.32     &   0.06    &  0.08     & 0.00  &  0.00     &  0.11   & 0.16 &   0.00    &   0.00    &      0.00    &   0.00 \\ 
YOLOWorld             & \checkmark          &  0.42     & 0.45      &    0.47   &  0.56     &  0.29     &    0.32   &   0.59    &   0.65    &  0.48     &   0.55    &   0.37    &   0.41    &   0.63    & 0.67 &    0.71   &    0.73   &    0.43   &    0.53   &   0.00    &  0.01    \\ 
MR-CNN R50   & \checkmark          &  0.50     &   0.54    &  0.67     &   0.72    &   0.40    &    0.43   &   0.68    &   0.73    &   0.66    &   0.71    &   0.49    &  0.52     &    0.63   &     0.70    &   0.71    &   0.75    &   0.66    & 0.70  & 0.05   &  0.09   \\ 
MR-CNN DINOv2      & \checkmark          &  0.41     &  0.44     &  0.49     &  0.55     &   0.34    &  0.36     &    0.57   &   0.62    &  0.45     &  0.51     &   0.41    &   0.44    &    0.55   & 0.60 &  0.60     &  0.63     &   0.51    &   0.57   &  0.12     &   0.13  \\ 
GDINO         &                     &   0.11    &    0.48   &   0.17    &  0.57     &   0.06    &    0.33   &    0.06   &   0.34    &   0.31    &   0.83    &   0.07    &    0.36   &    0.03   & 0.21  &   0.12    &   0.47    &  0.08     &   0.42    &    0.00   & 0.04   \\ 
GDINO         & \checkmark          &   \textbf{0.69}    &   \textbf{0.78}    &  \textbf{0.76}     &  \textbf{0.84}     &     \textbf{0.58}  &  \textbf{0.69}     &    \textbf{0.74}   &    \textbf{0.80}   &   \textbf{0.85}    &   \textbf{0.91}    &    \textbf{0.62}   &    \textbf{0.72}   &   \textbf{0.79}    & \textbf{0.83}  & \textbf{0.83}      &   \textbf{0.86}    &  \textbf{0.67}     &  \textbf{0.76}     &  \textbf{0.33}     & \textbf{0.54}   \\ 
\bottomrule
\end{tabular}
\end{table*}

Table~\ref{conditional_eval_table} presents the detection performance of the models under the conditions outlined in Section~\ref{conditional_evaluation}. Once again, GDINO FT (Fine-Tuned) achieved the highest performance across all evaluation criteria, consistently achieving the highest AP/AR values under different lighting conditions: 0.69/0.78 (WL), 0.76/0.84 (NBI); across the cancerous evolution: 0.58/0.69 (NICE 1), 0.74/0.8 (NICE 2), 0.85/0.91 (NICE 3); within smaller polyp sizes: 0.62/0.72 (Diminutive), 0.79/0.83 (Small); and even across various morphologies: 0.83/0.86 (Protruded), 0.67/0.76 (Flat Elevated), 0.33/0.54 (Flat). Notably, it was the only model to show meaningful detection within the flat polyps category.

MR-CNN R50 FT was the second-best performer, particularly in NBI Light Type (0.67/0.72) and NICE Type 2 (0.68/0.73). YOLOv8 (Fine-Tuned) showed competitive results in Protruded polyps (0.75/0.80), but struggled with flat classifications (0.59/0.64 at best).

In terms of the aforementioned robustness to light, even if no model has outperformed NBI with WL, the detection in WL with GDINO (0.69/0.78) is superior to any of the other models’ NBI performance; second best NBI results are 0.67/0.72 obtained with MR-CNN R50.

\section{Discussion}
\label{discussion}

\subsection{Discussion of General Performance}

For the detection task (Table \ref{detection_table}), fine-tuned GDINO achieved the highest values for both Average Precision (AP) and Average Recall (AR) across all datasets. These results demonstrate an effective detection of present polyps while minimizing false negatives; an important factor in colonoscopy, where missing a polyp has severe consequences. Also, the scores achieved by GDINO are 7.9\% to 26.5\% higher than those of its state-of-the-art competitor, MR-CNN R50, highlighting that the superior generalization ability of foundation models remains robust also in the medical domain. 

A great portion of this success can be attributed to the fine-tuning process, which allowed the foundation models to adapt effectively to the task. This impact was particularly evident in YOLO-World, where fine-tuning led to a remarkable improvement. Notably, baseline GDINO obtained AR values comparable to the best performing models. Its low AP, alongside a comparatively higher AR, suggests that, while the model generates a significant number of false positives, it still manages to capture a substantial portion of the relevant objects. With its distinct strategy of prioritizing quantity over precision, the model proved efficient in the context of assisted colonoscopy, where the objective is leaving no polyp behind.

Lastly, as for DINOv2, contrary to previous statements, the foundation model did not improve MR-CNN's detection when used as an alternative backbone to the traditional R50.

In terms of segmentation, fine-tuned GDINO paired with MedSAM consistently produced the best results across all datasets. The combination of GDINO's detection capabilities with MedSAM's segmentation expertise demonstrated exceptional efficacy in medical imaging, supporting the choice of an instance segmentation approach and a dual-model strategy for greater accuracy. 

MedSAM generally outperformed SAM, emphasizing the importance of domain-specific knowledge. While both models share the same architecture, MedSAM's fine-tuning for medical imaging allows it to handle biological tissues and other anatomic structures better compared to the general-purpose SAM model.

On dataset variability: SUN-SEG and PolypSegm-ASH were included to address the clinical representation gaps of PICCOLO. SUN-SEG accounts for no-polyp cases, while PolypSegm-ASH introduces variability with 473 unique polyps. Performance trends were consistent across datasets, though not strictly proportional. PolypSegm-ASH generally produced the highest scores, even surpassing PICCOLO, indicating good adaptation to diverse morphologies. SUN-SEG, on the other hand, proved more challenging, suggesting there is room to improve the algorithm’s false positive rate. However, given its intended role as a clinical aid, prioritizing no false negatives was considered beneficial.

\begin{figure}[t]
    \centering
    \begin{subfigure}{1.0\columnwidth} 
        \centerline{
        \includegraphics[width=\linewidth]{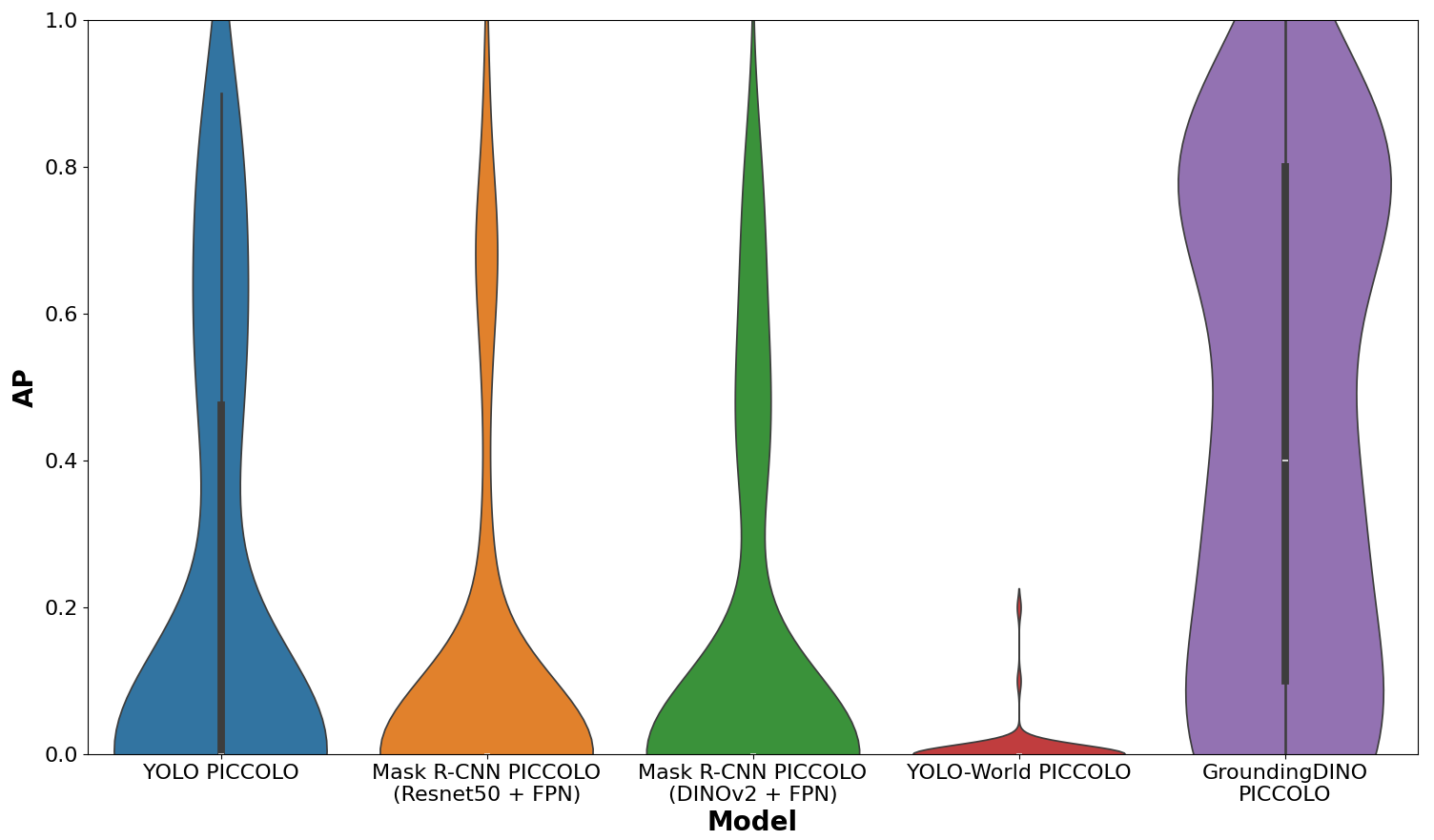}}
        \label{fig:subfig1}
    \end{subfigure}
    \caption{\textbf{Samplewise AP for Flat polyp detection (Paris classification)}. Left to right: YOLOv8 FT, MR-CNN R50 FT, MR-CNN DinoV2 FT, YOLOWorld FT, GDINO FT.}
    \label{fig:violin_flat}
\end{figure}

\subsection{Discussion of Conditional Evaluation}

\begin{figure*}[!t]
\centering

    \begin{tabular}{>{\centering\arraybackslash}m{55pt} >{\centering\arraybackslash}m{80pt} >{\centering\arraybackslash}m{80pt} >{\centering\arraybackslash}m{80pt} >{\centering\arraybackslash}m{80pt} >{\centering\arraybackslash}m{80pt}}
    Dataset & Input Image & MR-CNN FT & GDINO FT + MedSAM & GDINO + SAM & Ground Truth \\
    
    PICCOLO Protruded &
    \includegraphics[width=\linewidth]{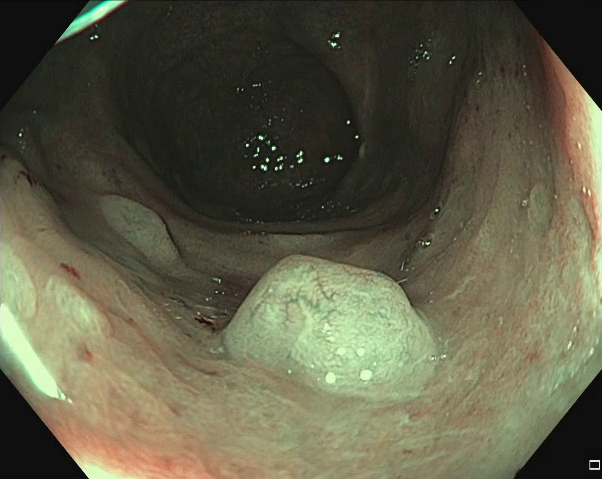} & 
    \includegraphics[width=\linewidth]{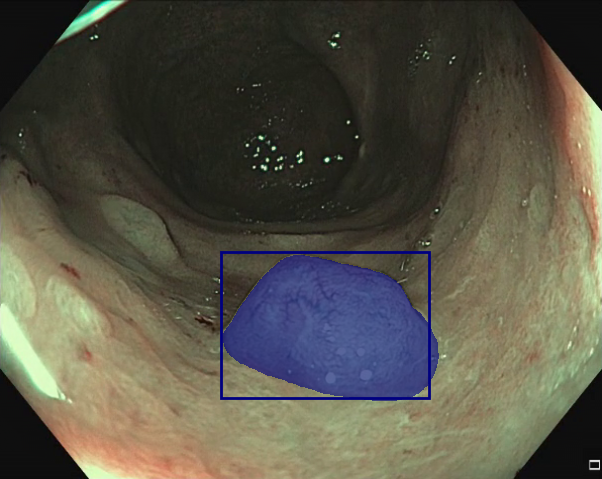} & 
    \includegraphics[width=\linewidth]{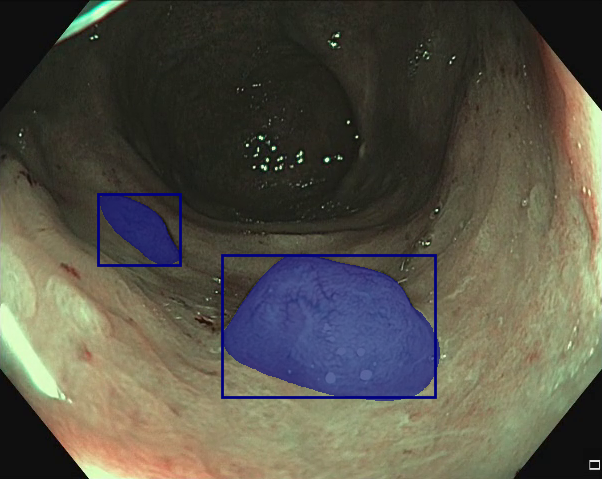} & 
    \includegraphics[width=\linewidth]{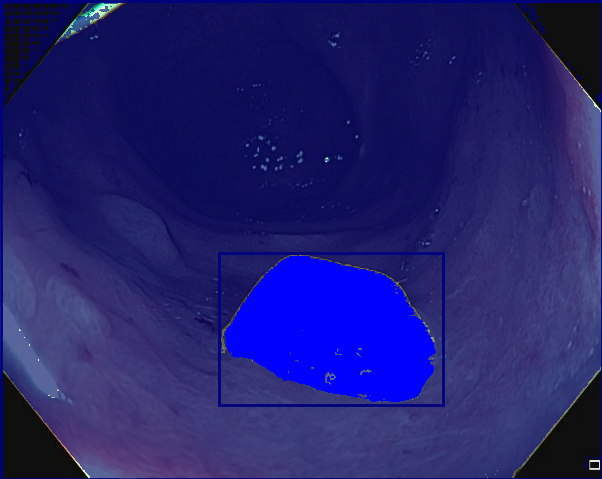} & 
    \includegraphics[width=\linewidth]{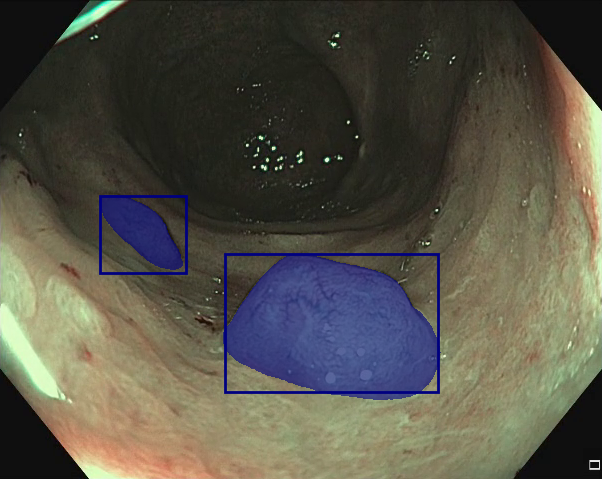} \\ 

    PICCOLO Flat elevated & 
    \includegraphics[width=\linewidth]{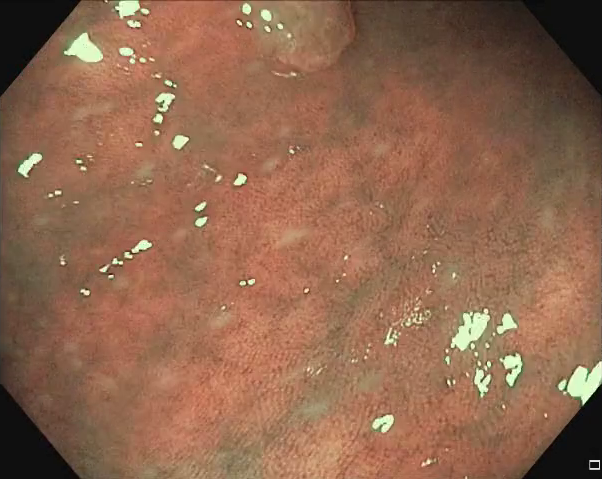} & 
    \includegraphics[width=\linewidth]{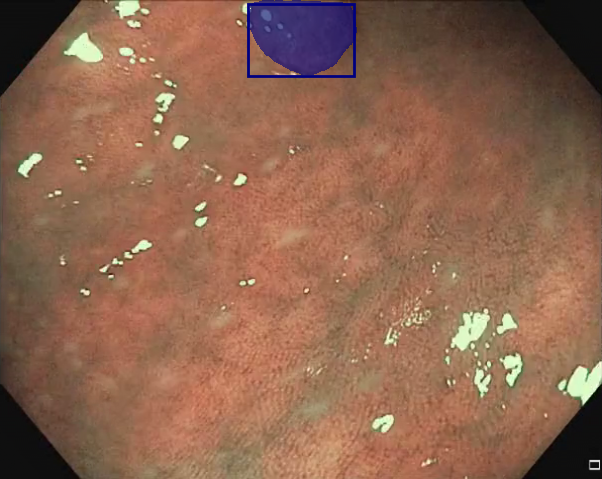} & 
    \includegraphics[width=\linewidth]{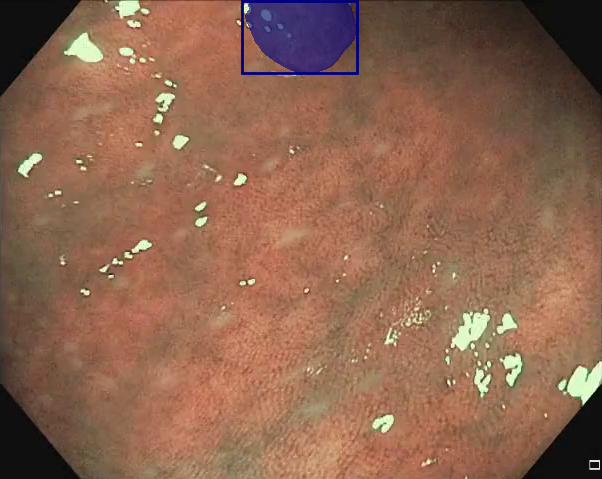} & 
    \includegraphics[width=\linewidth]{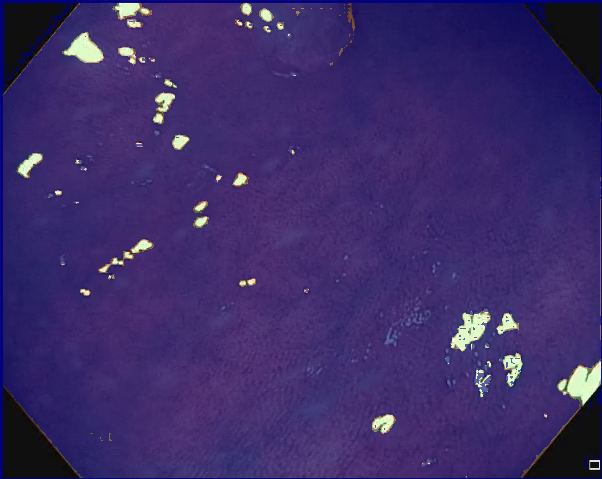} & 
    \includegraphics[width=\linewidth]{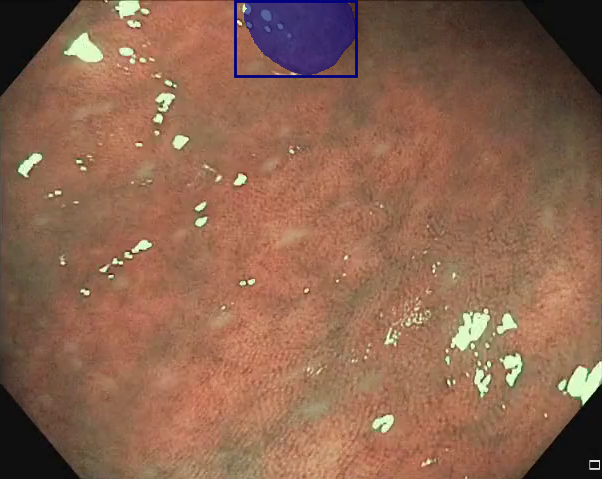} \\

    PICCOLO Flat & 
    \includegraphics[width=\linewidth]{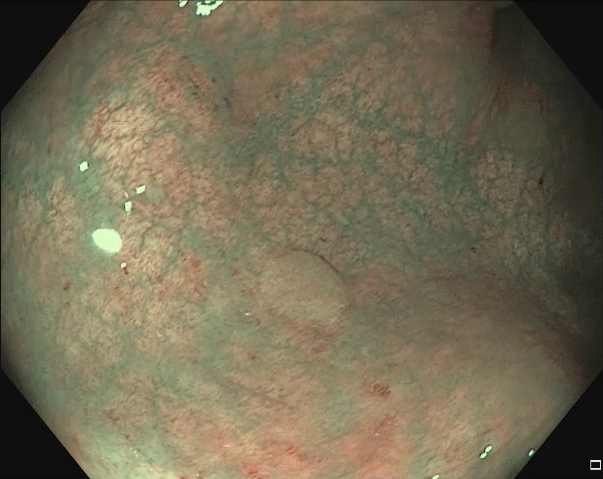} & 
    \includegraphics[width=\linewidth]{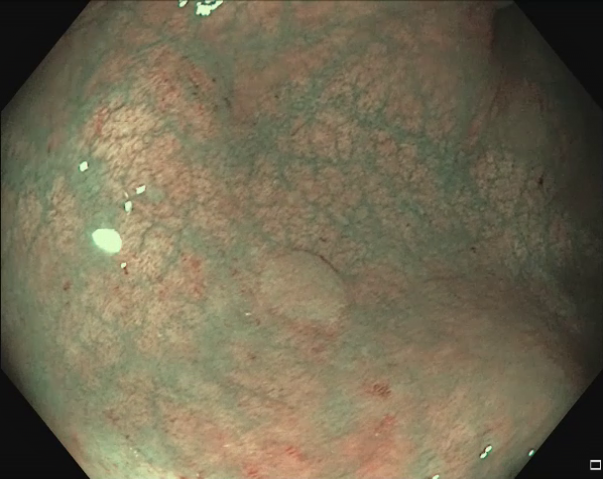} & 
    \includegraphics[width=\linewidth]{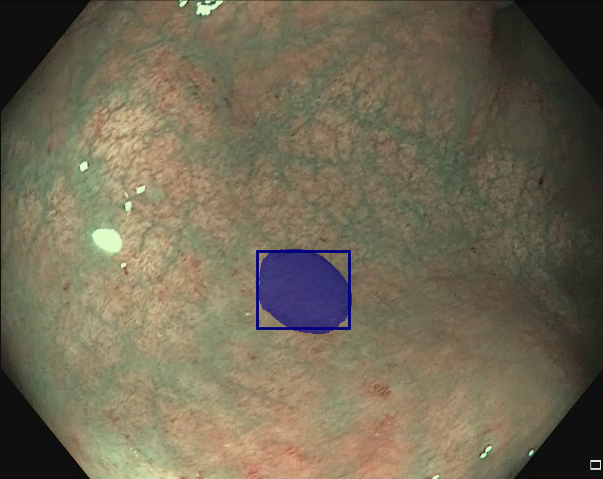} & 
    \includegraphics[width=\linewidth]{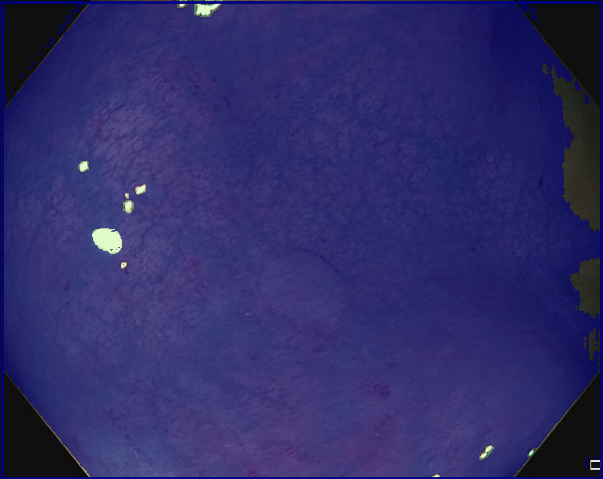} & 
    \includegraphics[width=\linewidth]{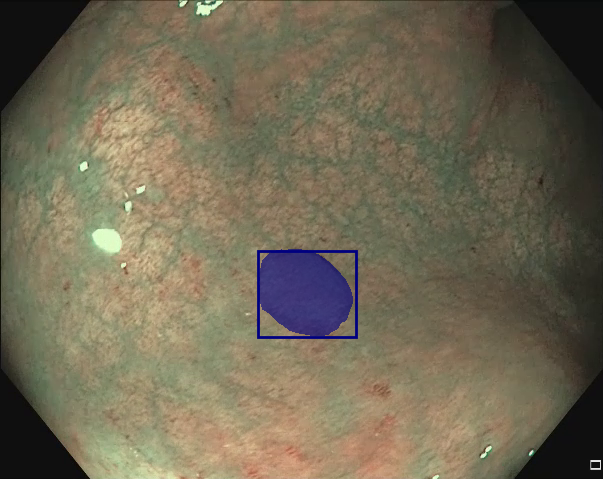} \\ 

    PolypSegm-ASH & 
    \includegraphics[width=\linewidth]{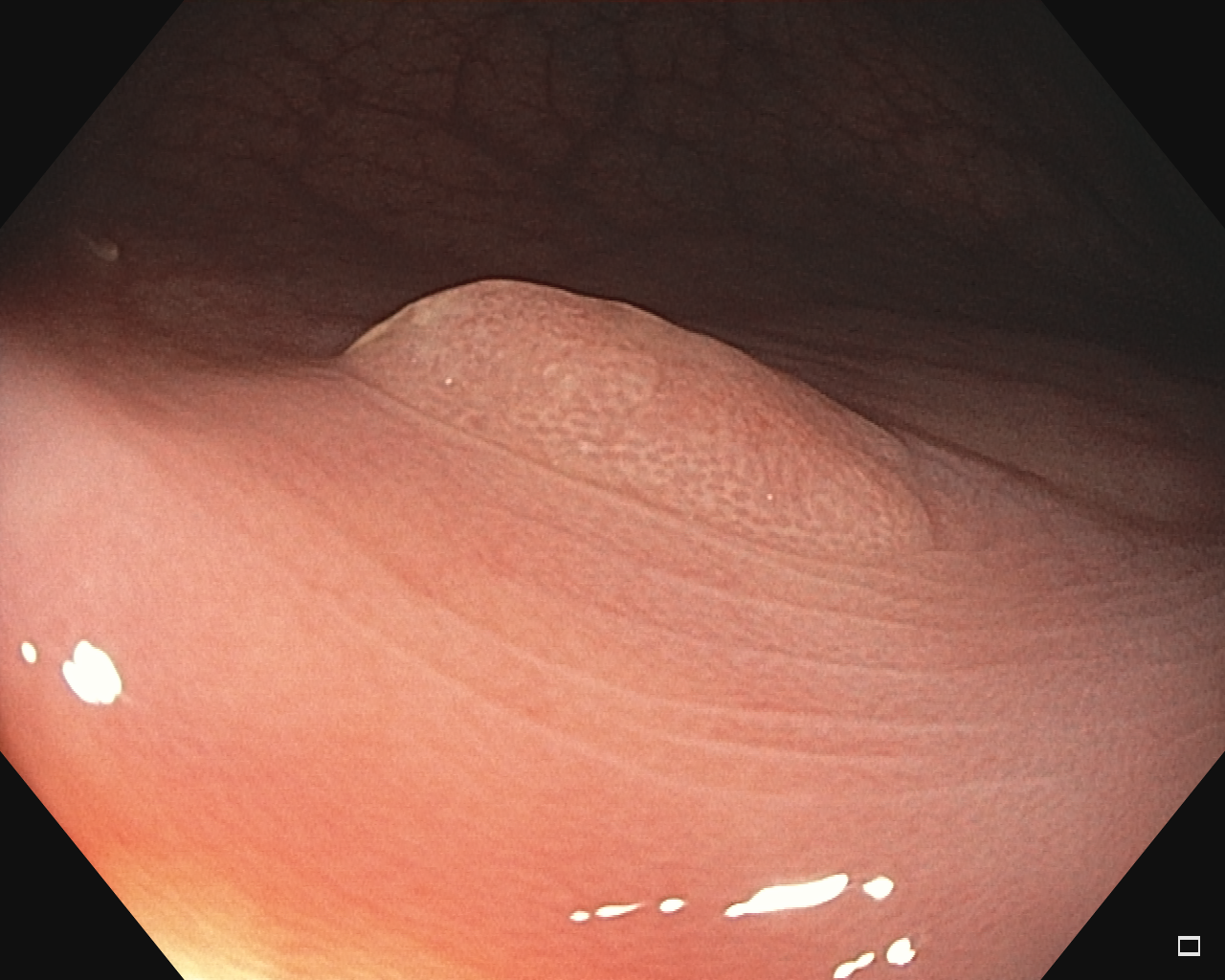} & 
    \includegraphics[width=\linewidth]{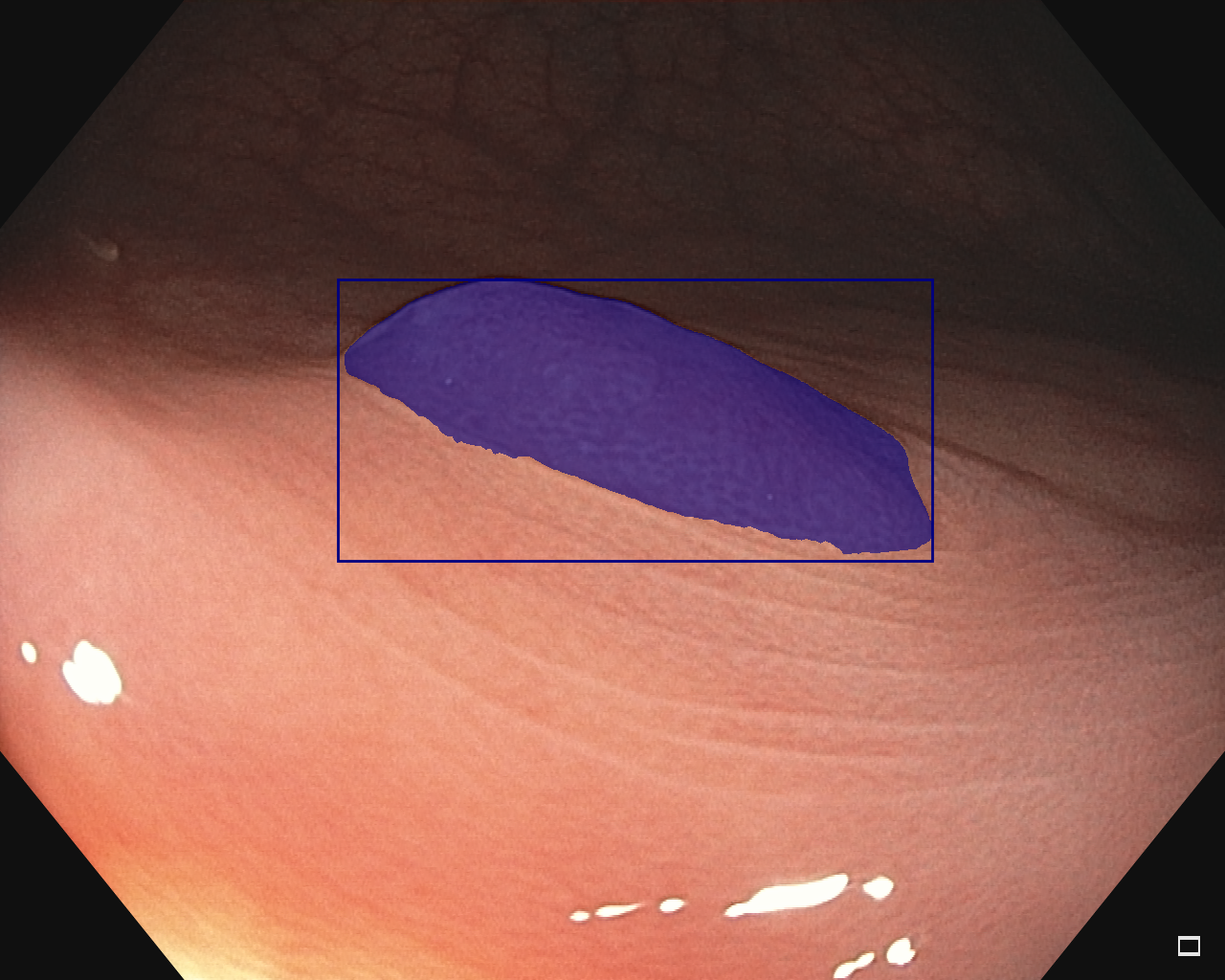} &
    \includegraphics[width=\linewidth]{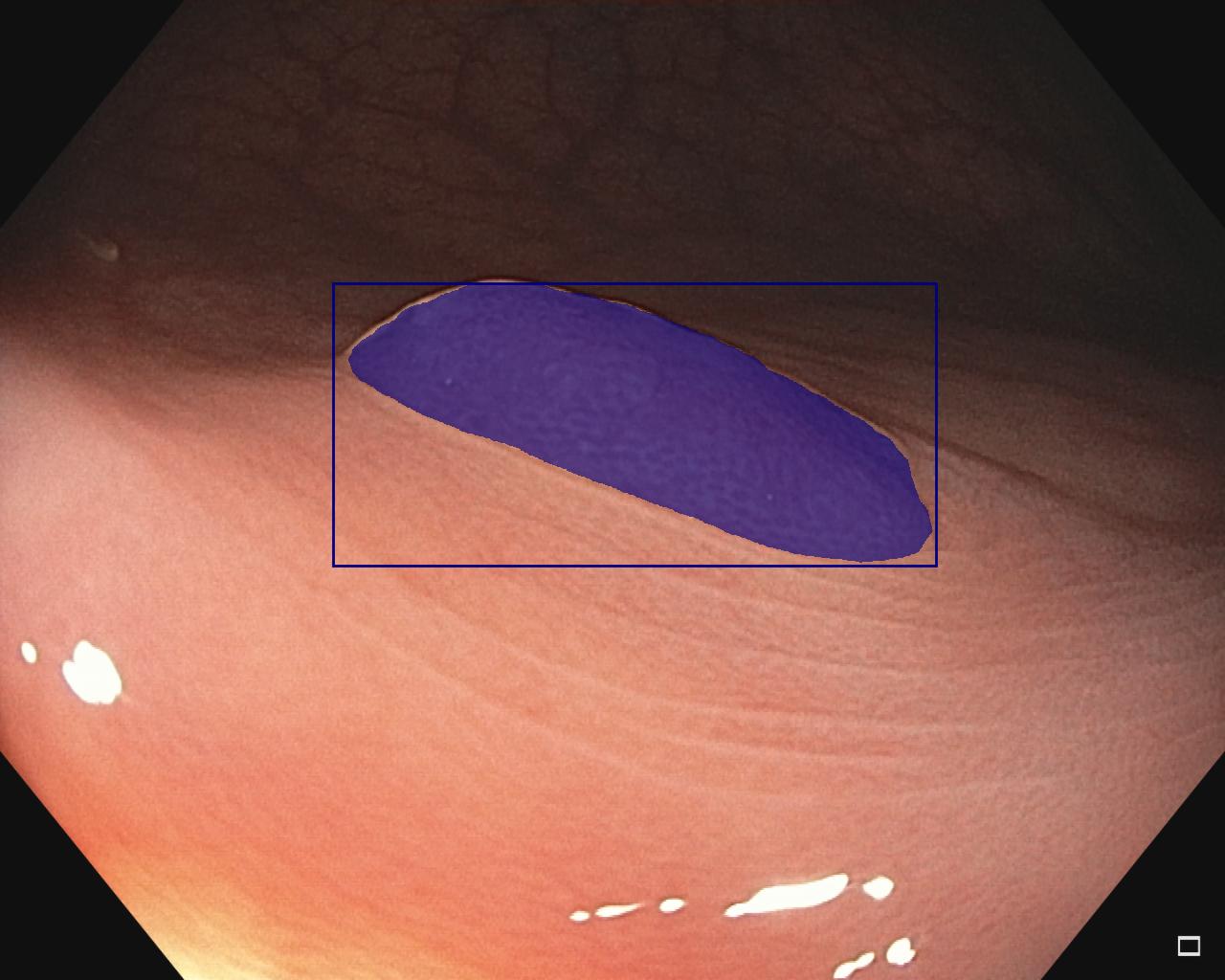} &  
    \includegraphics[width=\linewidth]{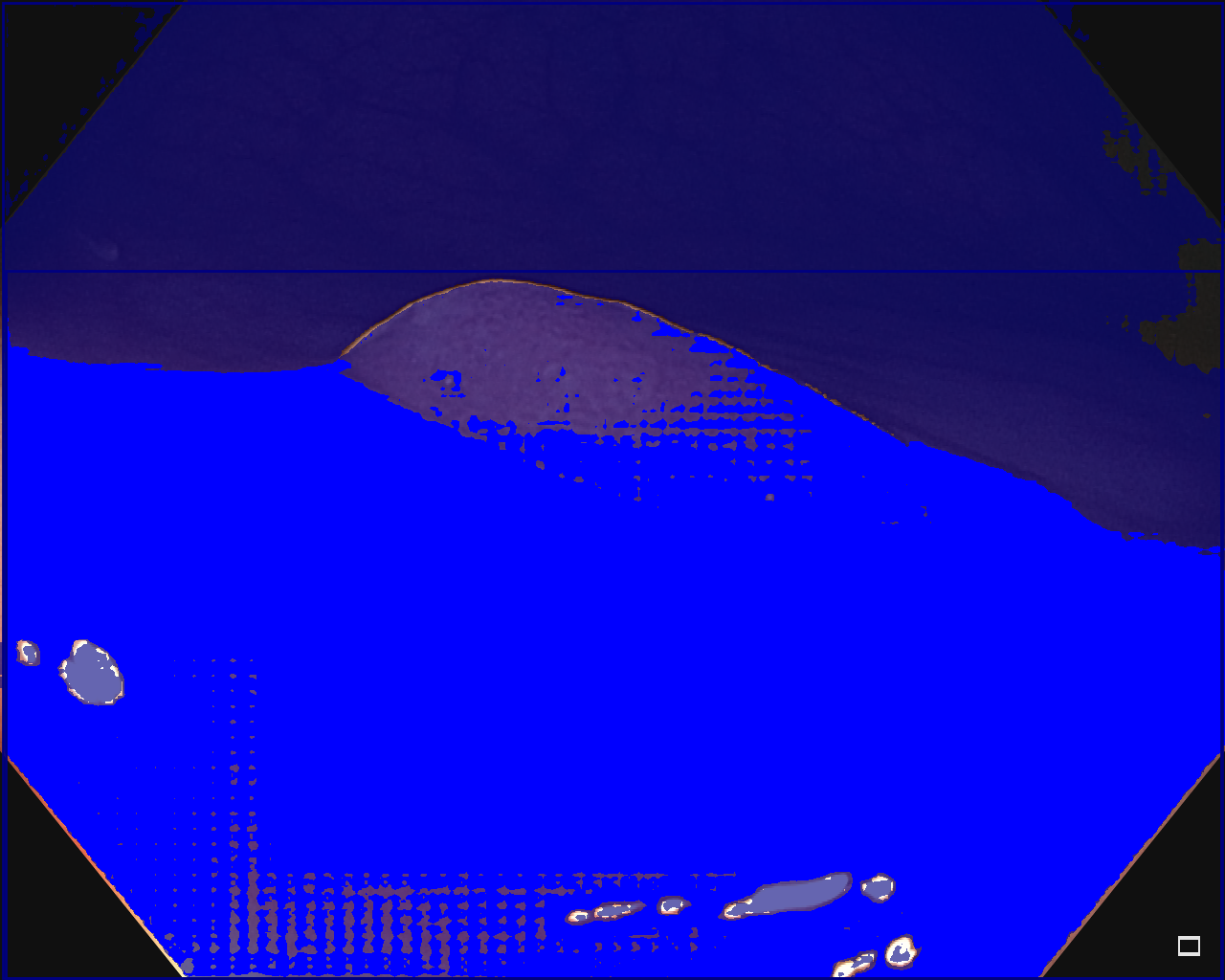} &
    \includegraphics[width=\linewidth]{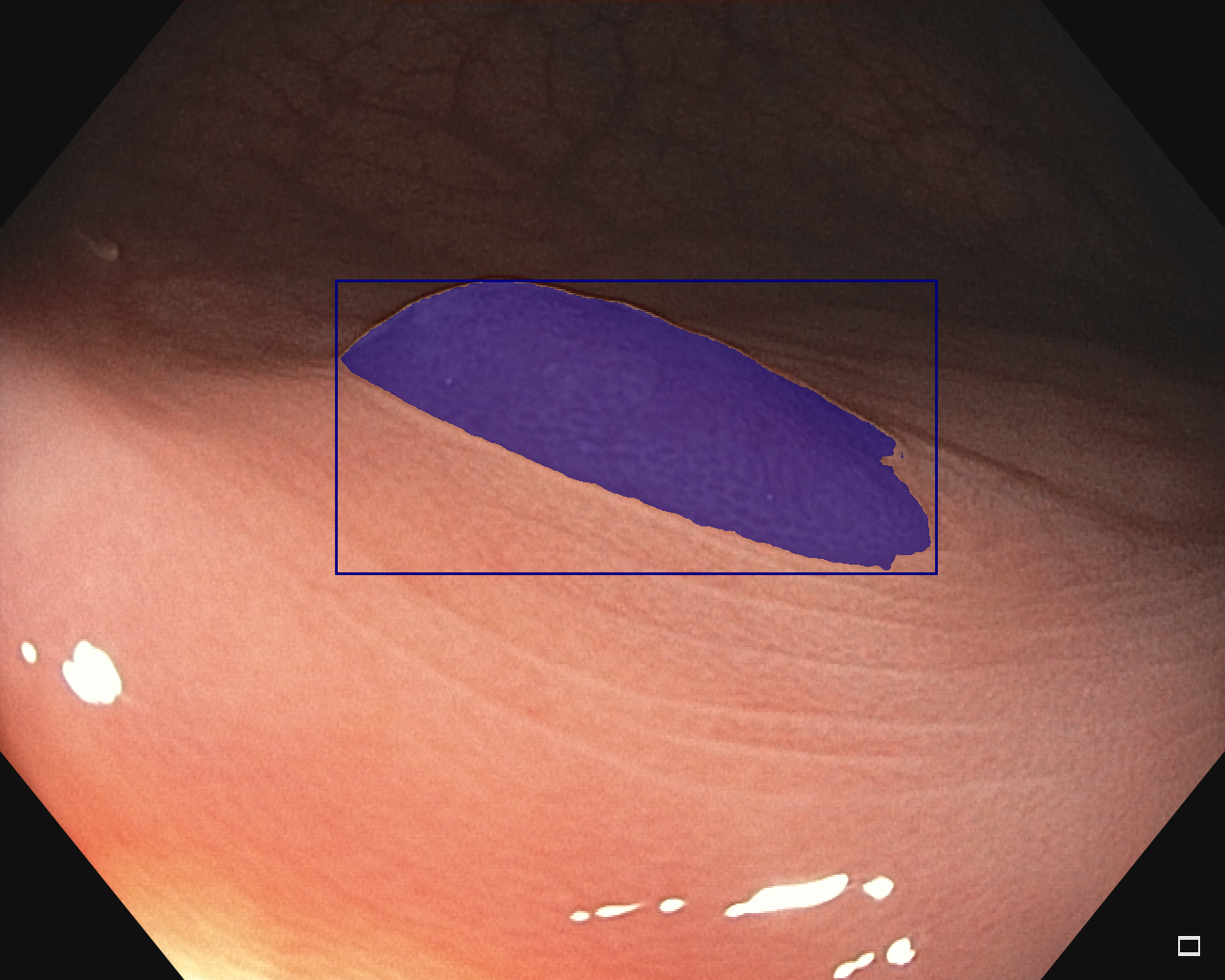} \\ 

    SUN-SEG & 
    \includegraphics[width=\linewidth]{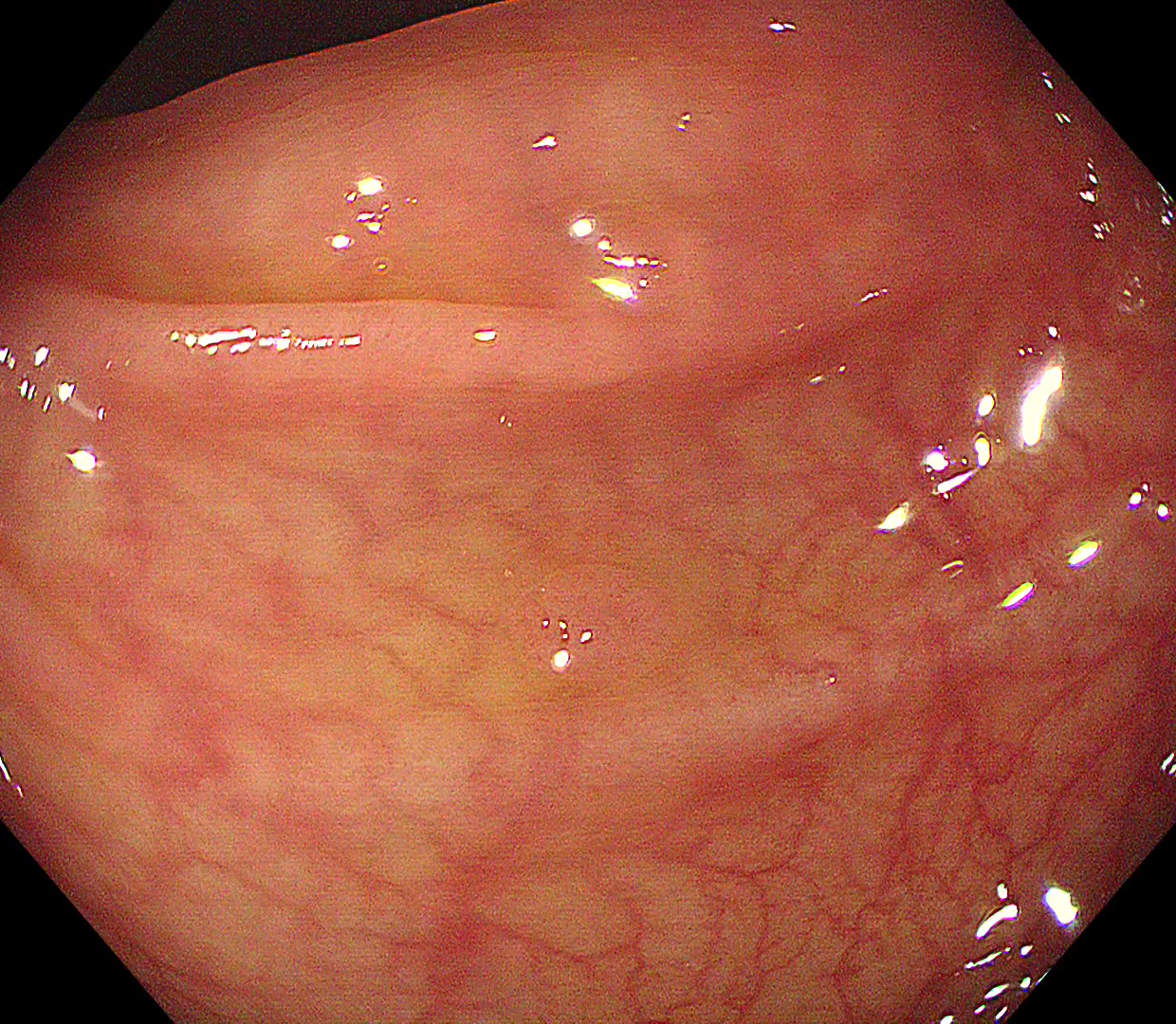} & 
   \includegraphics[width=\linewidth]{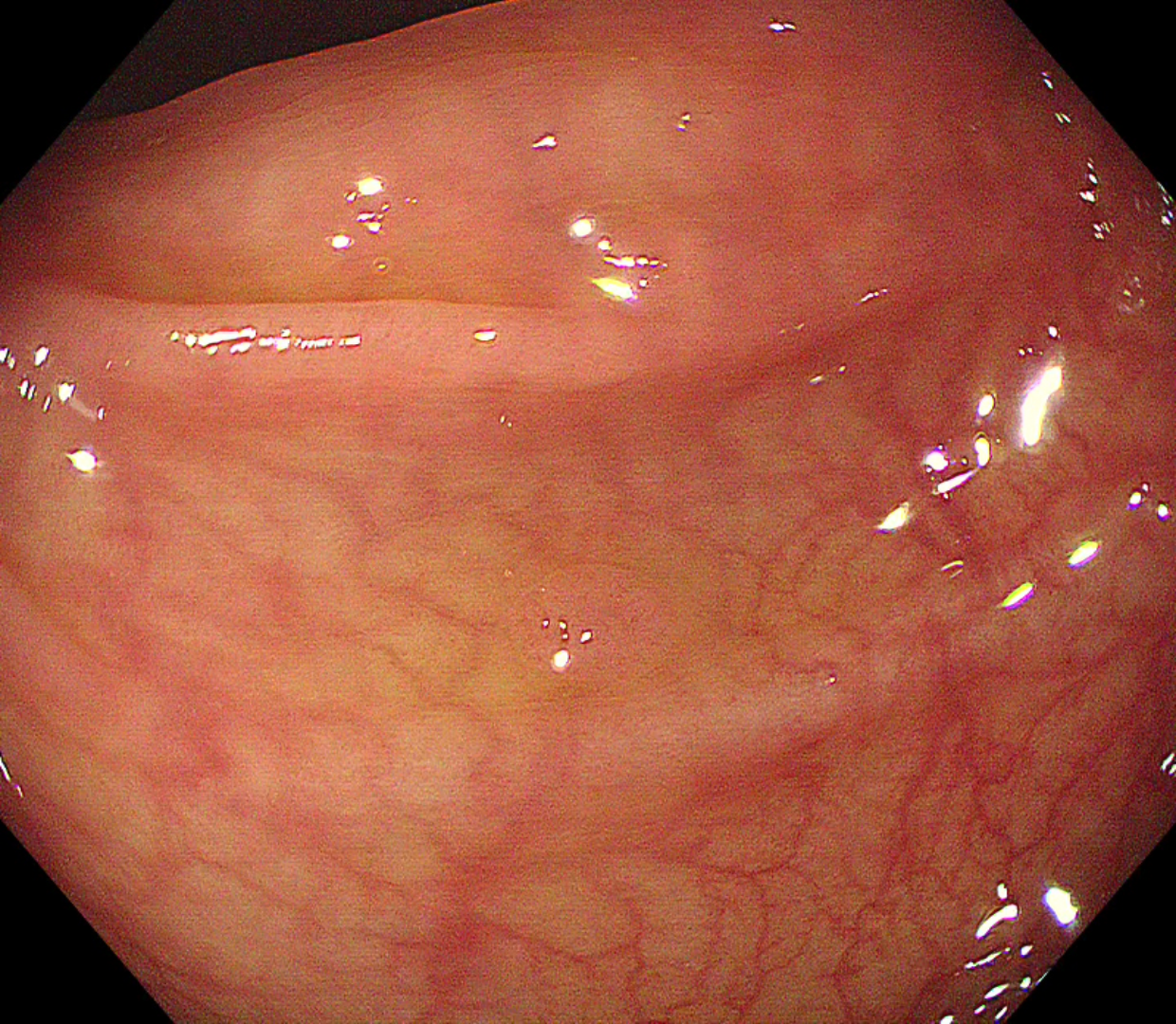} & 
    \includegraphics[width=\linewidth]{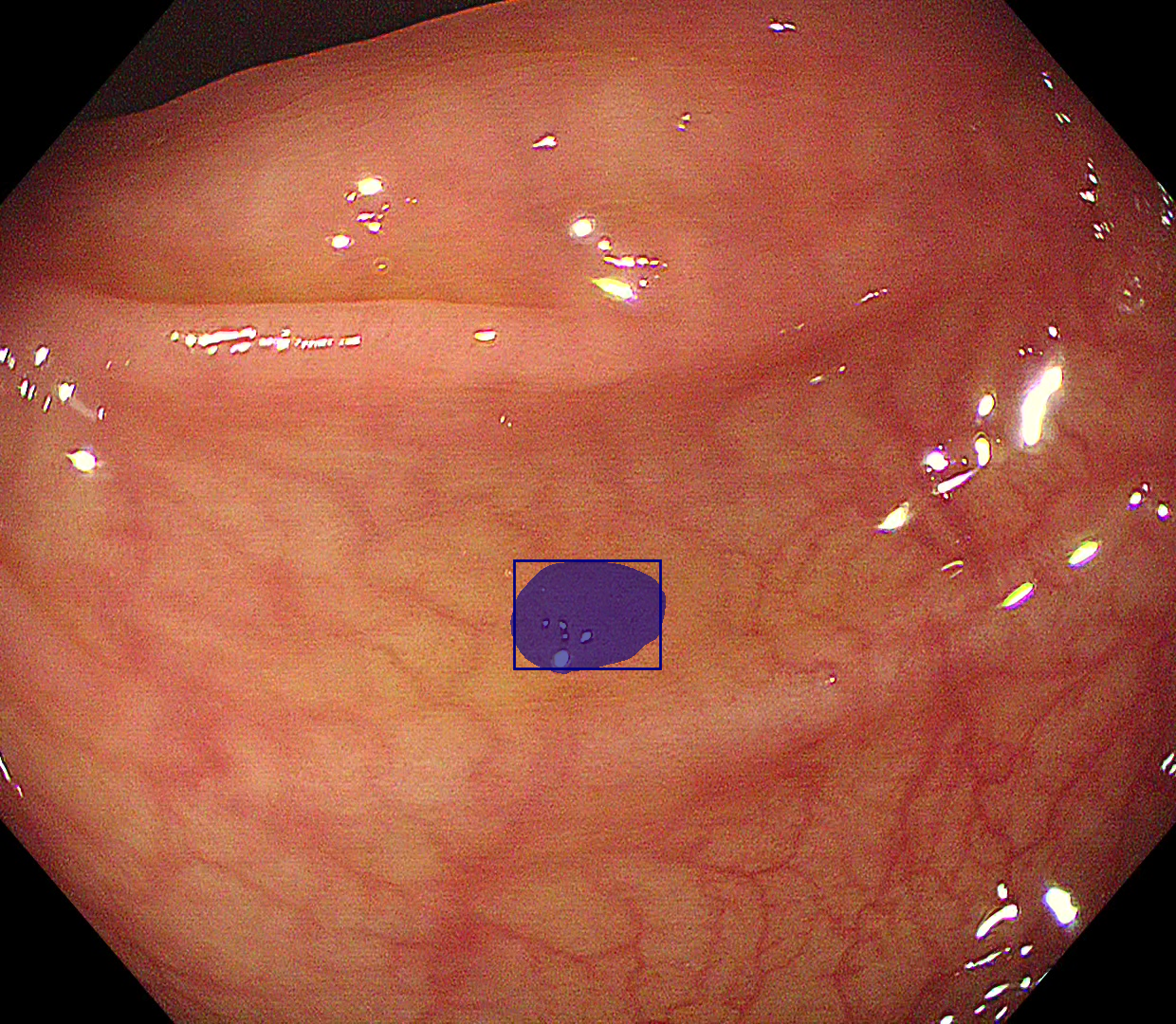} & 
    \includegraphics[width=\linewidth]{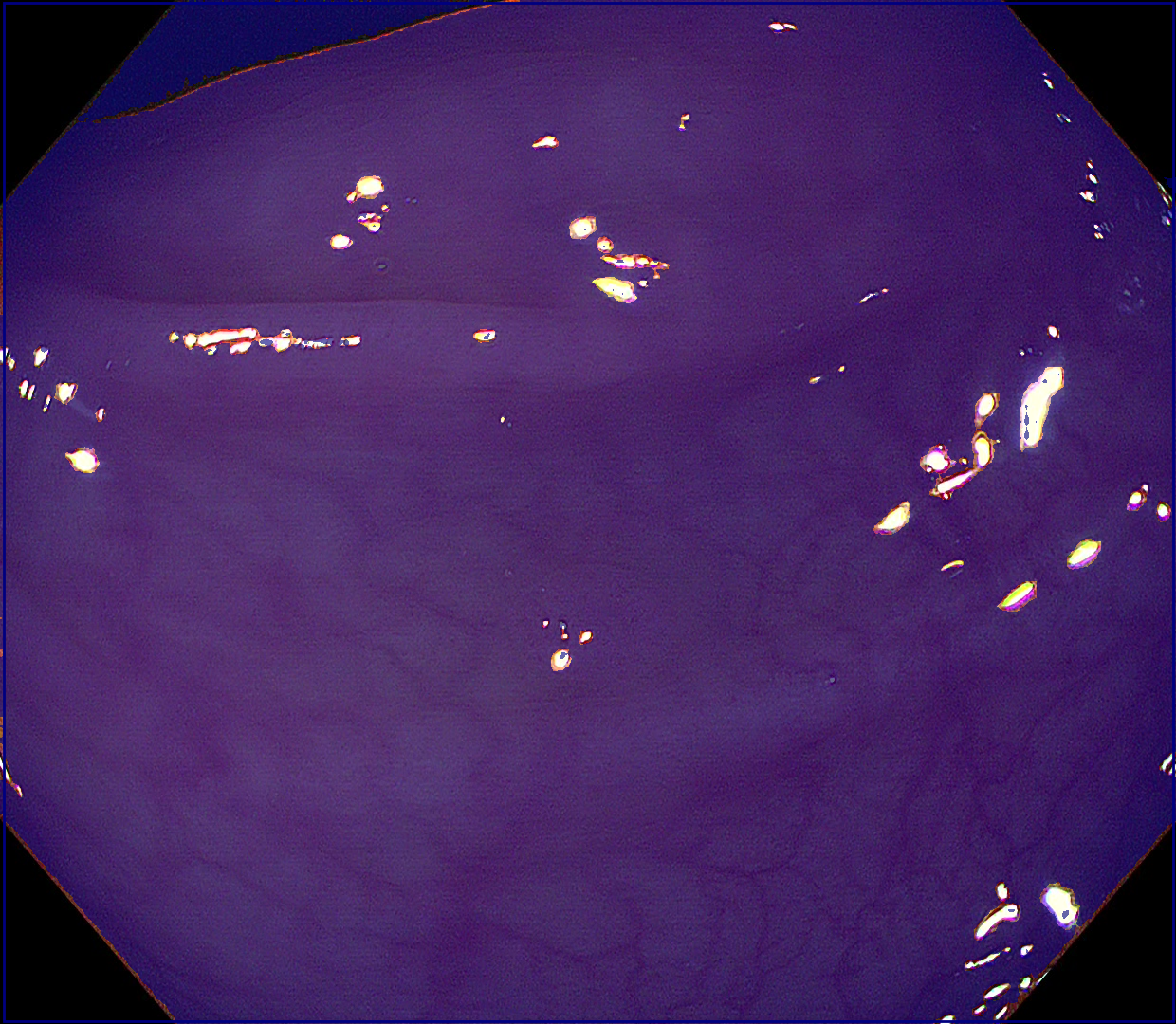} &
    \includegraphics[width=\linewidth]{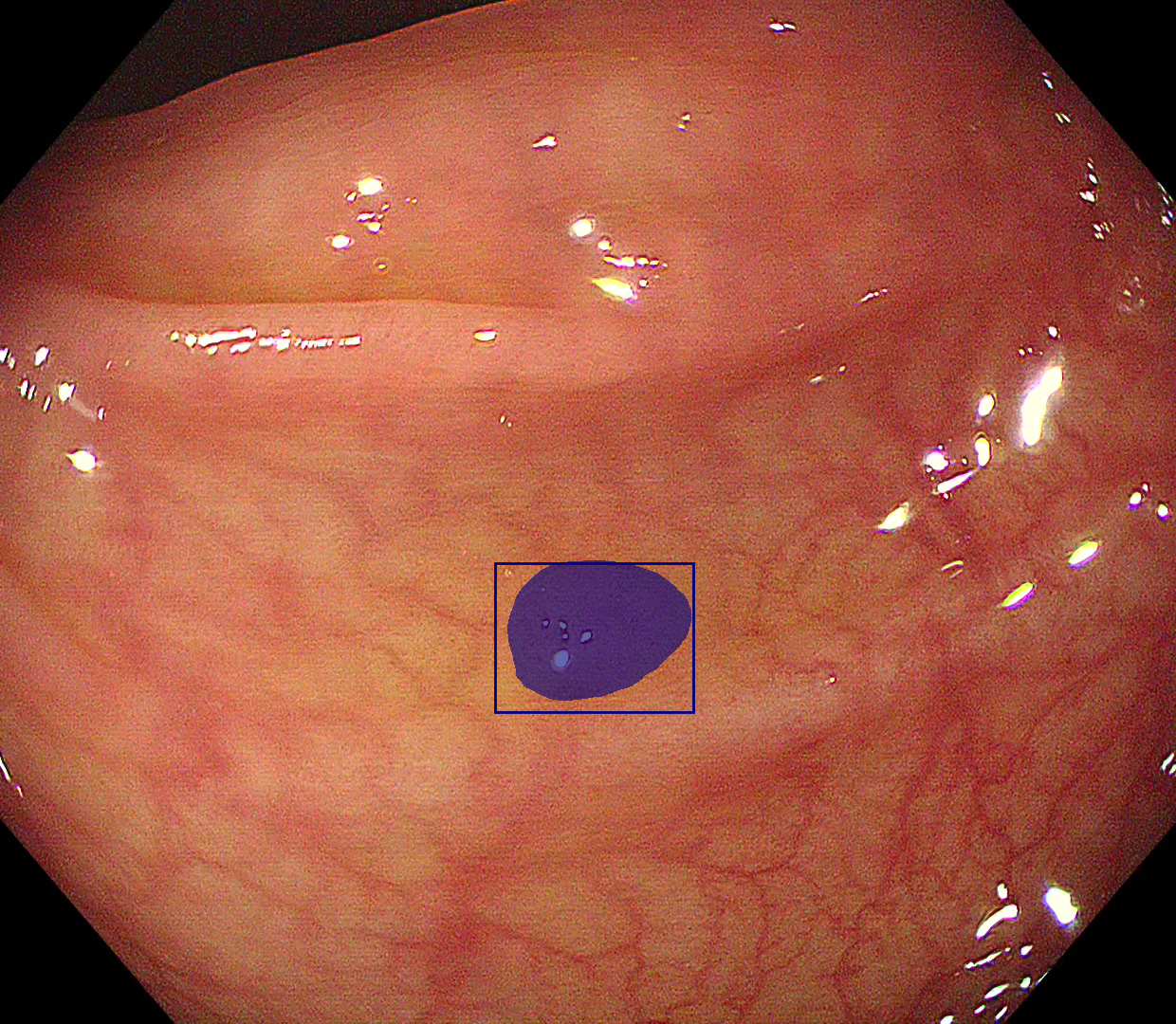} \\ 
    
    \end{tabular}
    \caption{\textbf{Performance on the analyzed images.} First column shows the original input image (from PICCOLO, PolypSegm-ASH or SUN-SEG datasets) with the corresponding Paris classification for PICCOLO. Second, third and fourth columns present the detection/segmentation results from the fine-tuned Mask R-CNN (baseline), fine-tuned GroundingDINO + MedSAM (best-performing) and GroundingDINO + SAM (worst-performing), respectively. Final column provides the ground truth annotations.}
    \label{fig: results_piccolo}
\end{figure*}

The conditional evaluation highlighted specific challenges associated with detecting polyps under different conditions based on their precision and recall scores. Polyps captured under white light (WL), those classified as type 1 NICE, diminutive polyps, and flat polyps were identified as the most difficult cases for detection. 

To further investigate these problematic cases, violin plots were employed to compare the density of samplewise detection metrics across the evaluated models.
In an ideal scenario, where polyps are detected with high accuracy, the plot would exhibit a wide distribution at the top and a narrow distribution at the bottom.
Among the evaluated models, GDINO demonstrated a distribution closest to this ideal in all difficult cases. The performance of this model was particularly notable for flat polyps (Fig. \ref{fig:violin_flat}), showing a significant increase in the samplewise detection AP/AR. 
The density of samplewise detections with an AP above 0.2 increased substantially, with the majority of cases clustering around the 0.8 AP mark, highlighting GDINO's superior performance. 

In addition to the quantitative analysis, the visual illustration of model predictions provides a deeper understanding of how these models would behave in a clinical setting. Fig.~\ref{fig: results_piccolo} compares the outputs of our best-performing model (Fine-tuned GDINO + MedSAM) against the worst-performing (GDINO + SAM) and the baseline (Fine-tuned Mask R-CNN) for samples drawn from the PICCOLO (for the three Paris categories), PolypSegm-ASH and SUN-SEG datasets.

The fine-tuned GDINO (FT GDINO + MedSAM) detected polyps in all the examples, including the more challenging flat and flat elevated morphologies. By observing the images, it becomes clear why flat polyps are particularly difficult to identify; their minimal elevation and visual similarity to healthy tissue make them easy to overlook. Yet, FT GDINO + MedSAM effectively captured them. Moreover, the best model could also recognize multiple instances within the same image, as seen in the first row where two polyps were individually detected. Row 5 shows a diminutive, almost imperceptible polyp that the baseline Mask R-CNN FT fails to detect but GDINO FT + MedSAM successfully identifies. In contrast, the non-fine-tuned model (GDINO + SAM) failed to produce any meaningful detections at all, reinforcing the necessity of fine-tuning for clinically relevant segmentation. \looseness=-1

\section{Conclusions}

Addressing the biggest question raised by the study, results suggest the need for domain-specialization of foundation models to excel in polyp detection and segmentation. For optimal performance in medical applications, specialized models are necessary, and generic models require fine-tuning to be effective, even with limited domain-specific data. This is likely due to the nature of their training data, which predominantly consists of everyday objects rather than medical structures.

MedSAM, as a domain-specific foundation model, demonstrates a clear advantage in understanding polyp characteristics. Remarkably good performance has been observed for MedSAM when fed with accurate detection prompts, performing better than fine-tuned models on data it has never seen. 

On a similar note, GDINO, after fine-tuning, achieved highly accurate detections across diverse scenarios, surpassing all competing models. Beyond overall detection effectiveness, recall (AR) plays a crucial role in ensuring reliability in clinical applications by minimizing false negatives. GDINO’s high AR values highlight its potential for real-world application, where missed detections can have serious consequences. 

Despite variations in dataset complexity, the best-performing model combination—fine-tuned GDINO with MedSAM—consistently demonstrated superior performance across all three datasets. This adaptability and robustness in diverse scenarios, reinforce its viability for clinical use. Moreover, the proposed detection and segmentation model maintains accuracy across challenging polyp categories and varying imaging conditions, making it a promising tool for enhancing polyp detection in colonoscopy. 

\section{Additional Information}
\label{additional}
Comprehensive details on dataset distribution criteria, prior model analyses, fine-tuning parameter settings and extended results are publicly vailable\footnote{\url{https://github.com/udelaqui/foundation_models_for_polyp_detection_segmentation}}.

\bibliographystyle{IEEEtran}
\bibliography{references}

\begin{thebibliography}{10}
\providecommand{\url}[1]{#1}
\csname url@samestyle\endcsname
\providecommand{\newblock}{\relax}
\providecommand{\bibinfo}[2]{#2}
\providecommand{\BIBentrySTDinterwordspacing}{\spaceskip=0pt\relax}
\providecommand{\BIBentryALTinterwordstretchfactor}{4}
\providecommand{\BIBentryALTinterwordspacing}{\spaceskip=\fontdimen2\font plus
\BIBentryALTinterwordstretchfactor\fontdimen3\font minus \fontdimen4\font\relax}
\providecommand{\BIBforeignlanguage}[2]{{%
\expandafter\ifx\csname l@#1\endcsname\relax
\typeout{** WARNING: IEEEtran.bst: No hyphenation pattern has been}%
\typeout{** loaded for the language `#1'. Using the pattern for}%
\typeout{** the default language instead.}%
\else
\language=\csname l@#1\endcsname
\fi
#2}}
\providecommand{\BIBdecl}{\relax}
\BIBdecl

\bibitem{CRC}
E.~J. Kuipers~et al., ``\BIBforeignlanguage{en}{Colorectal cancer},'' \emph{\BIBforeignlanguage{en}{Nature Reviews Disease Primers}}, vol.~1, no.~1, pp. 1--25, Nov. 2015.

\bibitem{acs_colorectal_2025}
A.~C. Society, ``Key statistics for colorectal cancer,'' \url{https://www.cancer.org/cancer/types/colon-rectal-cancer/about/key-statistics.html}, Dec. 2024.

\bibitem{ccstats}
R.~Siegel, N.~S. Wagle, A.~Cercek, R.~Smith, and A.~Jemal, ``Colorectal cancer statistics, 2023,'' \emph{CA: A Cancer Journal for Clinicians}, vol.~73, 03 2023.

\bibitem{cure}
W.~Atkin~et al., ``Once-only flexible sigmoidoscopy screening in prevention of colorectal cancer: A multicentre randomised controlled trial,'' \emph{Lancet}, vol. 375, pp. 1624--1633, May 2010.

\bibitem{origin}
I.~Mármol, C.~Sánchez~de Diego, A.~Pradilla-Dieste, E.~Cerrada, and M.~Rodríguez-Yoldi, ``Colorectal carcinoma: A general overview and future perspectives in colorectal cancer,'' \emph{Int. J. Mol. Sci.}, vol.~18, no.~1, p. 197, Jan. 2017.

\bibitem{malignantpolyp}
L.~Bujanda, A.~Cosme, I.~Gil, and J.~Arenas-Mirave, ``Malignant colorectal polyps,'' \emph{World J. Gastroenterol.}, vol.~16, no.~25, pp. 3103--3111, Jul. 2010.

\bibitem{piccoloproject}
J.~Ortega Morán~et al., ``Medical needs related to the endoscopic technology and colonoscopy for colorectal cancer diagnosis,'' \emph{BMC Cancer}, vol.~21, Apr 2021.

\bibitem{missedpolyps}
N.~Kim~et al., ``Miss rate of colorectal neoplastic polyps and risk factors for missed polyps in consecutive colonoscopies,'' \emph{Intestinal Research}, vol.~15, no.~3, p. 411, Jul 2017.

\bibitem{SOTA1}
L.~F. Sanchez-Peralta, L.~Bote-Curiel, A.~Picon, F.~Sánchez-Margallo, and J.~B. Pagador, ``Deep learning to find colorectal polyps in colonoscopy: A systematic literature review,'' \emph{Artificial Intelligence in Medicine}, vol. 108, p. 101923, Aug 2020.

\bibitem{foundationmodels}
A.~Narayan, I.~Chami, L.~Orr, and C.~R\'{e}, ``Can foundation models wrangle your data?'' \emph{Proc. VLDB Endow.}, p. 738–746, 12 2022.

\bibitem{SAM}
A.~Kirillov~et al., ``Segment anything,'' in \emph{Int. Conf. Comput. Vis. (ICCV)}, 2023.

\bibitem{MedSAM}
J.~Ma, Y.~He, F.~Li, L.~Han, C.~You, and B.~Wang, ``Segment anything in medical images,'' \emph{Nature Communications}, vol.~15, no.~1, p. 654, Jan. 2024.

\bibitem{dinov2}
M.~Oquab~et al., ``{DINO}v2: Learning robust visual features without supervision,'' \emph{Transactions on Machine Learning Research}, 2024.

\bibitem{yoloworld}
T.~Cheng, L.~Song, Y.~Ge, W.~Liu, X.~Wang, and Y.~Shan, ``Yolo-world: Real-time open-vocabulary object detection,'' in \emph{IEEE/CVF Conf. Comput. Vis. Pattern Recog. (CVPR)}, 2024.

\bibitem{groundingdino_paper}
S.~Liu~et al., ``Grounding dino: Marrying dino with grounded pre-training for open-set object detection,'' in \emph{Eur. Conf. Comput. Vis. (ECCV)}, 2024.

\bibitem{conference}
J.~P. Huix, A.~R. Ganeshan, J.~F. Haslum, M.~Soderberg, C.~Matsoukas, and K.~Smith, ``{ Are Natural Domain Foundation Models Useful for Medical Image Classification? },'' in \emph{IEEE/CVF Winter Conf. Appl. Comput. Vis. (WACV)}, 01 2024, pp. 7619--7628.

\bibitem{miccai}
J.~Bernal~et al., ``Comparative validation of polyp detection methods in video colonoscopy: Results from the miccai 2015 endoscopic vision challenge,'' \emph{IEEE Trans. Med. Imag.}, vol.~36, no.~6, pp. 1231--1249, Feb 2017.

\bibitem{urban2018}
G.~Urban~et al., ``Deep learning localizes and identifies polyps in real time with 96\% accuracy in screening colonoscopy,'' \emph{Gastroenterology}, vol. 155, 06 2018.

\bibitem{maskrcnn}
K.~He, G.~Gkioxari, P.~Dollar, and R.~Girshick, ``Mask r-cnn,'' \emph{IEEE Trans. Pattern Anal. Mach. Intell.}, vol.~PP, pp. 1--1, 06 2018.

\bibitem{chenbaldi}
S.~Chen, G.~Urban, and P.~Baldi, ``Weakly supervised polyp segmentation in colonoscopy images using deep neural networks,'' \emph{Journal of Imaging}, vol.~8, p. 121, 04 2022.

\bibitem{alamfattah}
M.~Alam and S.~A. Fattah, ``{SR-AttNet}: An interpretable stretch–relax attention based deep neural network for polyp segmentation in colonoscopy images,'' \emph{Computers in Biology and Medicine}, vol. 160, p. 106945, 04 2023.

\bibitem{chen}
\BIBentryALTinterwordspacing
J.~Chen~et al., ``{3D TransUNet}: Advancing medical image segmentation through vision transformers,'' 2023. [Online]. Available: \url{https://arxiv.org/abs/2310.07781}
\BIBentrySTDinterwordspacing

\bibitem{sameg}
Q.-H. Trinh, H.-D. Nguyen, N.~N.~B. Tram, D.~Jha, U.~Bagci, and M.-T. Tran, ``Sam-eg: Segment anything model with edge guidance framework for efficient polyp segmentation,'' in \emph{Brit. Mach. Vis. Conf. (BMVC)}.\hskip 1em plus 0.5em minus 0.4em\relax BMVA, 2024.

\bibitem{polypsam}
\BIBentryALTinterwordspacing
R.~Biswas, ``Polyp-{SAM}++: {Can} {A} {Text} {Guided} {SAM} {Perform} {Better} for {Polyp} {Segmentation}?'' Aug. 2023. [Online]. Available: \url{https://arxiv.org/abs/2308.06623}
\BIBentrySTDinterwordspacing

\bibitem{yolo}
J.~Redmon, S.~Divvala, R.~Girshick, and A.~Farhadi, ``You only look once: Unified, real-time object detection,'' in \emph{IEEE Conf. Comput. Vis. Pattern Recog. (CVPR)}.\hskip 1em plus 0.5em minus 0.4em\relax {IEEE}, 2016, pp. 779--788.

\bibitem{clip}
A.~Radford~et al., ``Learning transferable visual models from natural language supervision,'' in \emph{Int. Conf. Machi. Learn. (ICML)}, 2021.

\bibitem{dino}
H.~Zhang~et al., ``{DINO}: {DETR} with improved denoising anchor boxes for end-to-end object detection,'' in \emph{Int. Conf. Learn. Represent. (ICLR)}, 2023.

\bibitem{bert}
J.~Devlin, M.-W. Chang, K.~Lee, and K.~Toutanova, ``Bert: Pre-training of deep bidirectional transformers for language understanding,'' in \emph{Proc. NAACL}, 2019.

\bibitem{PICCOLO}
L.~F. Sanchez-Peralta~et al., ``Piccolo white-light and narrow-band imaging colonoscopic dataset: A performance comparative of models and datasets,'' \emph{Applied Sciences}, vol.~10, no.~12, p. 8501, Dec. 2020.

\bibitem{Paris}
A.~Axon~et al., ``Update on the paris classification of superficial neoplastic lesions in the digestive tract,'' \emph{Endoscopy}, vol.~37, pp. 570--578, Jun. 2005.

\bibitem{nice}
S.~Hattori~et al., ``Narrow-band imaging observation of colorectal lesions using nice classification to avoid discarding significant lesions,'' \emph{World J. Gastrointest. Endosc.}, vol.~6, no.~12, pp. 600--605, Dec. 2014.

\bibitem{polypsegm-ash}
Y.~Tudela, A.~Garc{\'i}a-Rodr{\'i}guez, G.~Fern{\'a}ndez-Esparrach, and J.~Bernal, ``Towards fine-grained polyp segmentation and classification,'' in \emph{Clinical Image-Based Procedures, Fairness of AI in Medical Imaging, and Ethical and Philosophical Issues in Medical Imaging}.\hskip 1em plus 0.5em minus 0.4em\relax Springer Nature Switzerland, 2023, pp. 32--42.

\bibitem{sunseg}
M.~Misawa~et al., ``Development of a computer-aided detection system for colonoscopy and a publicly accessible large colonoscopy video database (with video),'' \emph{Gastrointestinal Endoscopy}, vol.~93, 07 2020.

\bibitem{yolov8_doc}
G.~Jocher, A.~Chaurasia, and J.~Qiu, ``Ultralytics yolov8 documentation,'' \url{https://docs.ultralytics.com/models/yolov8/}, 2024, accessed: 2024-06-09.

\bibitem{yoloworld_doc}
\BIBentryALTinterwordspacing
{Ultralytics}, ``Yolo-world models,'' 2024. [Online]. Available: \url{https://docs.ultralytics.com/models/yolo-world/}
\BIBentrySTDinterwordspacing

\bibitem{groundingdino_mmdetection_ft}
\BIBentryALTinterwordspacing
K.~Chen~et al., ``{MMDetection}: Open mmlab detection toolbox and benchmark,'' 2019. [Online]. Available: \url{https://arxiv.org/abs/1906.07155}
\BIBentrySTDinterwordspacing

\bibitem{coco}
T.-Y. Lin~et al., ``Microsoft coco: Common objects in context,'' in \emph{Eur. Conf. Comput. Vis. (ECCV)}, 2014.

\end{thebibliography}

\end{document}